\documentclass[aps,prc,a4paper,nofootinbib,showpacs,showkeys,
preprintnumbers,superscriptaddress,twocolumn]{revtex4}

\usepackage{amsmath}
\usepackage{amssymb}
\usepackage{bm}
\usepackage{graphicx}
\usepackage{color}

\begin{document}

\title{Thermalization, Isotropization and Elliptic Flow 
from Nonequilibrium Initial Conditions with a Saturation Scale}

\author{M. Ruggieri}
\affiliation{Department of Physics and Astronomy, University of Catania, Via S. Sofia 64, I-95125 Catania}

\author{F. Scardina}
\affiliation{Department of Physics and Astronomy, University of Catania, Via S. Sofia 64, I-95125 Catania}
\affiliation{INFN-Laboratori Nazionali del Sud, Via S. Sofia 62, I-95123 Catania, Italy}

\author{S. Plumari}
\affiliation{Department of Physics and Astronomy, University of Catania, Via S. Sofia 64, I-95125 Catania}
\affiliation{INFN-Laboratori Nazionali del Sud, Via S. Sofia 62, I-95123 Catania, Italy}

\author{V. Greco}
\affiliation{Department of Physics and Astronomy, University of Catania, Via S. Sofia 64, I-95125 Catania}
\affiliation{INFN-Laboratori Nazionali del Sud, Via S. Sofia 62, I-95123 Catania, Italy}


\begin{abstract}
In this article we report on our results about the computation of
the elliptic flow of the quark-gluon-plasma produced in relativistic heavy ion
collisions, simulating the expansion of the fireball by solving
the relativistic Boltzmann equation for the parton distribution function tuned at a fixed shear viscosity to entropy density
ratio $\eta/s$. 
Our main goal is to put emphasis on the role of a saturation scale in the initial
gluon spectrum, which makes the initial distribution far from a thermalized one.
We find that the presence of the saturation scale reduces the efficiency
in building-up the elliptic flow, even if the thermalization process is quite fast $\tau_{therm} \approx 0.8 \,\rm fm/c$
and the pressure isotropization even faster $\tau_{isotr} \approx 0.5 \,\rm fm/c$.
The impact of the non-equilibrium implied by the saturation scale manifests for non-central collisions 
and can modify the estimate of the viscosity respect to the assumption of full thermalization in $p_T$-space.
We find that the estimate of $\eta/s$ is modified from $\eta/s \approx 2/4\pi$ to $\eta/s \approx 1/4\pi$  at RHIC and 
from  $\eta/s \approx 3/4\pi$ to $\eta/s \approx 2/4\pi$ at LHC.
We complete our investigation by a study of the thermalization and isotropization times of the
fireball for different initial conditions and values of $\eta/s$ showing how the latter affects 
both isotropization and thermalization. Lastly, we have seen that the range of values explored by the phase-space
distribution function $f$ is such that at $p_T<0.5\, \rm GeV$ the inner part of the fireball stays with occupation number
significantly larger than unity despite the fast longitudinal expansion, which might suggest the possibility
of the formation of a transient Bose-Einstein Condensate.

\end{abstract}


\maketitle

\section{Introduction}
In the last decade it has been reached a general consensus that Ultra-relativistic heavy-ion collisions (uRHICs) 
at the Relativistic Heavy-Ion Collider (RHIC) and the Large Hadron Collider (LHC) create a hot and dense 
strongly interacting quark and gluon plasma (QGP)   \cite{STAR_PHENIX, ALICE_2011, Science_Muller,Fries:2008hs}. 
A main discovery has been that the QGP has a very small shear viscosity to density entropy, $\eta/s$,
which is more than one order of magnitude smaller than the one of water~\cite{Csernai:2006zz,Lacey:2006bc}, 
and close to the lower bound of $1/4\pi$ conjectured for systems at infinite strong coupling \cite{Kovtun:2004de}. 
A key observable to reach such a conclusion is the elliptic flow~\cite{Ollitrault},
\begin{equation}
v_2 = \langle cos(2 \varphi_p) \rangle= 
\left\langle \frac{p_x^2-p_y^2}{p_x^2+p_y^2} \right\rangle~,
\end{equation}
with $\varphi_p$ being the azimuthal angle
in the transverse plane and the average meant over the particle distribution. 
In fact, the expansion of the created matter
generates a large anisotropy of the emitted particles that can be primarily measured by $v_2$.  
An hadronic system would generate a $v_2$ about a factor 2-3 smaller than the measured one \cite{Konchakovski:2011qa}
and this is one of the hints that the system created at RHIC and LHC is of non-hadronic nature.

The origin of $v_2$ is the initial spatial eccentricity, 
\begin{equation}
\epsilon_x=
 \frac{\langle y^2-x^2\rangle}{\langle x^2+y^2 \rangle}~,
\end{equation} 
of the overlap region in non-central collisions, which is responsible for different
pressure gradients in the transverse plane thus favoring flow preferably 
along the $x$ direction rather than  $y$ direction.
The observed large $v_2$ is considered a signal of a very small $\eta/s$
because it means that the system is very efficient in converting $\epsilon_x$
into an anisotropy in the momentum space $v_2$, a mechanism that would be strongly damped in a  
system highly viscous that dissipates and smooths anisotropies \cite{Romatschke:2007mq, Heinz,Cifarelli:2012zz}.
Quantitatively both viscous hydrodynamics \cite{Romatschke:2007mq,Heinz,Song:2011hk,Schenke:2010nt,Niemi:2011ix}, 
and  transport Boltzmann-like 
approaches~\cite{Ferini:2008he, Xu:2008av,Xu:2007jv,Bratkovskaya:2011wp,Plumari_BARI,Ruggieri:2013bda}
agree in indicating an average $\eta/s$ of the QGP lying in the range $4\pi\eta/s \sim 1-3$.

Along with the existence of a deconfined QGP matter and the understanding of its properties, 
the uRHIC program offers the opportunity to verify the picture in which at vey high energy
the two colliding nuclei are described as two sheets of
Color Glass Condensate (CGC)~\cite{McLerran:1993ni,McLerran:1993ka,McLerran:1994vd},
see~\cite{Gelis:2010nm,Iancu:2003xm,McLerran:2008es,Gelis:2012ri} for reviews.
The CGC in the nuclei would be primarily generated by the very high density of the gluon distribution
function at low $x$ (parton momentum fraction), which triggers a saturation
of the distribution for $p_T$ below a saturation scale, $Q_s$. 
The determination of the shear viscosity $\eta/s$ of the QGP and the search for the CGC are related.
In fact, the main source of uncertainty for $\eta/s$ comes from the unknown initial conditions 
of the created matter~\cite{Luzum:2008cw, Alver:2010dn,Song:2011hk,Adare:2011tg} that imply
quite different eccentricities $\epsilon_x$.  

A simple geometrical description through the Glauber model~\cite{Miller:2007ri} predicts a 
$\epsilon_x$ smaller at least 25-30$\%$ than the eccentricity of the CGC, for most of the centralities of 
the collisions, see for example results within the
Kharzeev-Levin-Nardi (KLN) model~\cite{Kharzeev:2004if,Hirano:2005xf,Drescher:2006pi}, 
factorized KLN (fKLN) model~\cite{Drescher:2006ca}, 
Monte Carlo KLN (MC-KLN) model ~\cite{Drescher:2006ca,Hirano:2009ah} and dipole 
model~\cite{Drescher:2006pi,Albacete:2010pg}.  
In fact, the saturation of gluon distribution function is stronger in the central overlapping
region, making the space distribution sharper than the one coming from a simple
geometrical overlap. Other CGC models like the ones based on the solution of the classic Yang-Mills (CYM)
equations predict a somewhat smaller initial eccentricity~\cite{Lappi:2006xc}.
However the effect we discuss in the present paper is mainly related to the saturation effect in momentum
space and not in the $r-$space that determines the eccentricity, even if certainly to consider
CYM initial conditions is an important task considering the success it has in predicting
the collective flow anisotropies $v_n=\langle cos(n\phi)\rangle$ with $n\ge 2$ \cite{Gale:2012rq}.

In this article, we report our results about the computation of
the elliptic flow of the quark-gluon-plasma produced in relativistic heavy ion
collisions, simulating the expansion of the fireball by solving
the relativistic Boltzmann equation for the parton distribution function tuned at a fixed shear viscosity to entropy density
ratio $\eta/s$~\cite{Ferini:2008he,Plumari:2010ah,Plumari_BARI,Plumari:2012xz}. 
In this context the advantage of using kinetic theory is that starting from a one-body
phase space distribution function $f(x,p)$, and not from the energy-momentum tensor $T^{\mu\nu}$, 
it is straightforward to initialize simulations from a non-equilibrium distribution function, 
like the one characterizing the KLN model, while the hydrodynamics relies on the gradient expansion of the
stress tensor and therefore is applicable only if initial deviations from
local equilibrium are small. It is worth noticing that kinetic theory has also attracted
interest as a tool to derive equations for the hydrodynamical evolution
of a fluid starting from a microscopic theory~\cite{Tsumura:2013uma,Bazow:2013ifa}.

In this study we assume that the initial condition in configuration and momentum space arising from
the melting of the CGC is given by the KLN model.
In fact the KLN initial condition that has been largely employed to study
the dynamics of HIC and the viscosity of the QGP \cite{Song:2011hk,Luzum:2008cw,Alver:2010dn,Hirano:2009ah,
Shen:2013vja,Shen:2013cca,Song:2013qma}. 
The uncertainty in the initial condition 
translates into an uncertainty on $\eta/s$ of at a least a factor of two as estimated
by mean of several viscous hydrodynamical approaches \cite{Luzum:2008cw,Alver:2010dn,Song:2011hk,Adare:2011tg}. 
More explicitly, the experimental
data of $v_2(p_T)$ at the highest RHIC energy are in agreement with a fluid with $4\pi\eta/s\approx 1$
according to viscous hydrodynamics simulation, assuming a standard Glauber
initial condition. Assuming an initial  fKLN or MC-KLN space distribution the comparison
favors a fluid at $4\pi\eta/s\approx 2$. The reason is the larger initial $\epsilon_x$ 
of the fKLN, which leads to larger $v_2$ unless a large $\eta/s$ is considered.
We point out that the implementation of the shattered CGC initialization
in hydrodynamics takes into account only the different space distribution
respect to a geometric Glauber model,
discarding the key and more peculiar feature of the damping
of the distribution for $p_T$ below the $Q_s$ saturation scale.
We have found by mean of kinetic theory
that this has a significant impact on the build-up of $v_2$.

In order to frame our work correctly, we refer to the standard picture
of the evolution of the fireball produced in the heavy ion collision,
so we can identify the moment in which our approach is valid.
The commonly accepted picture goes as follows: at $\tau=0^-$ the
two high energy colliding nuclei are described as two thin sheets of
color-glass condensate, which after the collision turns at $\tau=0^+$ into 
an out-of-equilibrium glasma state, which consists of a background of chromomagnetic
and chromoelectric flux tubes, on the top of which quantum fluctuations
are produced and cause the decay of the glasma itself to a partong liquid
as the system expands. This parton system then equilibrates and the
evolution is governed by viscous hydro.
During the very early evolution of the glasma,
the initial negative longitudinal pressure becomes positive because
of the breaking of the color strings, see for 
example~\cite{Ryblewski:2013eja,Gelis:2013rba,Fukushima:2013dma}.
The decay causes the fields become very weak within a fraction
of fm/c: we denote this time by $\tau_0$. 
From this time a description of the system as a pure interacting partons 
liquid becomes reliable, and can be described by kinetic theory. 
We therefore start our simulations from $\tau_0$. 
We assume  $\tau_0 \sim Q_s^{-1}= 0.2$ fm/c in this work;
however we have checked that our main results on the elliptic flow
are not affected significantly by shifting the initial time to reasonably larger times.
Within our framework the longitudinal pressure
$P_L$ at $\tau=\tau_0$ is positive, in agreement with recent calculations 
within Classic-Yang Mills statistical approach \cite{Gelis:2013rba}
or within kinetic theory at fixed $\eta/s$ with a gauge field in abelian dominance \cite{Ryblewski:2013eja}.
Our main purpose
in this study is to take into account strong deviations from local equilibrium
in the initial condition, which naturally arise when the KLN model
is used in the calculations: these kind of deviations are not affordably treated within viscous
hydrodynamics even if in principle non vanishing bulk and shear energy tensor at $\tau_0$
may account for it. Anyway, we will show that our description
reproduces the hydrodynamics results when a proper thermalized initial condition
is prepared in our simulations.

The plan of the article is as follows. In Section II, we review the initial conditions
we use in the simulations, namely the factorized glasma the Glauber initial conditions.
In Section III, we review our approach to kinetic theory at fixed $\eta/s$. 
In Section IV, we present our results on thermalization and isotropization of the
fireball, as well as on the time evolution
of the parton distribution function in the expanding system. 
In Section V, we present and discuss our results on the
elliptic flow. Finally in Section VI, we draw our conclusions.

\section{Gluon production in the KLN approach}
In order to prepare the initial condition in coordinate and momentum space
of a melted glasma state we 
adopt the model which was firstly introduced
by Kharzeev, Levin and Nardi~\cite{Kharzeev:2004if}
(KLN model), with particular reference to
the factorized-KLN (fKLN) approach as introduced in~\cite{Drescher:2006ca,Hirano:2009ah}, 
in which the coordinate space distribution function of gluons arising from the melted glasma is assumed to be 
\begin{eqnarray}
\frac{d N_g}{dy d^2\bm{x}_\perp} = \int d^2 \bm{p}_T\, p_A(\bm x_\perp) p_B(\bm x_\perp)  
\frac{dN}{d{\cal P}} ~,
\label{eq:densityCS}
\end{eqnarray}
where $dN/d{\cal P}$ corresponds to the momentum space distribution in the $k_T$ factorization 
hypothesis~\cite{Gribov:1984tu,Kovchegov:2001sc},
\begin{eqnarray}
\frac{dN}{d{\cal P}} &=&\frac{4\pi^2 N_c}{N_c^2-1} \frac{1}{p_T^2}
\int^{p_T} d^2\bm{k}_T
\alpha_S(Q^2)\nonumber\\
&&\times\phi_A(x_1,k_T^2;\bm{x}_\perp)\nonumber\\
&&\times\phi_B(x_2,(\bm{p}_T -\bm{k}_T)^2;\bm{x}_\perp)~,
\label{eq:densityPS}
\end{eqnarray}
and $d{\cal P}=dy d^2\bm x_T d^2 \bm p_T$. Here $x_{1,2} = p_T \exp(\pm y)/\sqrt{s}$
correspond to the longitudinal momentum fraction carried by the gluons belonging
to the two colliding nuclei which produce a gluon with transverse momentum $p_T$ and
rapidity $y$; 
the ultraviolet cutoff in the $k_T$ integral in Eq.~\eqref{eq:densityCS} is $p_T = 3$ GeV$/c$  
for the case of Au-Au collisions at $\sqrt{s}=200 A$ GeV, while we take 
$p_T = 4$ GeV$/c$ for the case of Pb-Pb collisions at $\sqrt{s}=2.76 A$ TeV; 
$\alpha_S$ denotes the strong coupling constant,
which is computed at the scale $Q^2 = \text{max}(\bm k_T^2,(\bm p_T - \bm k_T)^2)$ according to
the one-loop $\beta$ function but frozen at $\alpha_s=0.5$ in the infrared region
as in~\cite{ALbacete:2010ad,Hirano:2005xf,Albacete:2010pg}.
In Eq.~\eqref{eq:densityCS} $p_{A,B}$ denote the probability to find one nucleon
at a given transverse coordinate, namely
\begin{equation}
p_A({\bm x}_\perp) = 1-
\left(1-{\sigma_{in}} T_A(\bm x_\perp)/A\right)^A~,
\label{eq:pA}
\end{equation}
where $\sigma_{in}$ is the inelastic cross section and  $T_A$ corresponds
to the usual thickness function of the Glauber model. 

Equation~\eqref{eq:densityCS} is based on the
factorization hypothesis, which is known to work fairly in the case
of proton-proton as well as proton-nucleus collisions but it is violated in the case of a
nucleus-nucleus collision~\cite{Blaizot:2010kh}. However, we do not aim at
considering the exact spectrum of a glasma, that may 
be computed for example within the CYM approach~\cite{Blaizot:2010kh,Lappi:2011ju}; 
rather we wish to study another issue, that is how the initial
nonequilibrium distribution in momentum space, in particular one with a
saturation scale built in, affects the building up of the
elliptic flow. Furthermore, the KLN initial condition is still
largely employed in hydrodynamics simulations 
to test the impact of initial conditions on different observables
\cite{Shen:2013vja,Shen:2013cca,Song:2013qma,Heinz:2013th,Moreland:2012qw,
Albacete:2011fw,Qiu:2011iv,Song:2011qa,Monnai:2011ju,Hirano:2010je}.
For simplicity, and also to uniform our language to
that commonly adopted in hydro simulations, we still refer to the
initial condition specified by Eq.~\eqref{eq:densityCS} as to the
melted-glasma initial condition.

The main ingredient to specify in Eq.~\eqref{eq:densityPS}  
is the unintegrated gluon distribution function (uGDF) for partons 
coming from nucleus $A$, which is assumed to be:
\begin{eqnarray}
\phi_A(x_1,k_T^2;\bm{x}_\perp) = \frac{\kappa \,Q_s^2}{\alpha_s(Q_s^2)}
\left[\frac{\theta(Q_s - k_T)}{Q_s^2 + \Lambda^2} + \frac{\theta(k_T - Q_s)}{k_T^2 + \Lambda^2}\right]~,
\label{eq:phiA}
\end{eqnarray}
which embeds saturation of the distribution
for $p_T< Q_s$; a similar equation holds for partons belonging to nucleus $B$. 
Following~\cite{Drescher:2006ca} we take the saturation scale for the nucleus $A$ as
\begin{equation}
Q_{s,A}^2(x,{\bm x}_\perp)=
Q_{0}^2\left(\frac{T_A({\bm x}_\perp)}{1.53 p_A({\bm x}_\perp)}\right)
\left(\frac{0.01}{x}\right)^\lambda~,
\label{eq:PPPppp}
\end{equation}
with $\lambda=0.28$, and similarly nucleus $B$. 
The scale $Q_{0}$ regulates the average value of the saturation scale
on the transverse plane. Equation \eqref{eq:PPPppp} implies that
for a fixed value of $x$, the larger the nucleon density the larger the saturation scale,
since $Q_s^2 \propto T_A$ for large $T_A$;
however in the limit $T_A \rightarrow 0$, which is realized in the peripheral region of  each
of the two colliding nuclei, Equation \eqref{eq:PPPppp} implies that
$Q_s$ does not vanish as $T_A$. In fact if $T_A\rightarrow 0$ then from Eq. \eqref{eq:pA} we get 
$T_A/p_A \rightarrow 1/\sigma_{in}$, which implies that $Q_s$ is nonzero in this limit,
and coincides with the saturation scale of a nucleon~\cite{Drescher:2006ca}.

For our simulations of Au-Au collisions at RHIC energy we take $Q_0^2=1$ GeV$^2$, 
which gives $\langle Q_s\rangle \approx 1$ GeV
in the case $b=0$, at $y=0$, $\sqrt{s}=200A$ GeV and $x=0.01$, where the average is understood
in the transverse plane. This numerical value is smaller than the one commonly used
in hydrodynamic simulations, where $\langle Q_s\rangle\approx 1.4$ GeV, see for 
example~\cite{Song:2011hk,Drescher:2006ca,Hirano:2009ah},
and that we used in our previous work~\cite{Ruggieri:2013bda}.
We prefer to use a smaller value of the average saturation scale
in our simulations because in this way we can shift from the fKLN
initial condition to the other ones obtaining the
same final spectra, without the need to tweak the kinetic freezout
parameters when changing among the several initial conditions.
In this way, the differences between the various systems we study here
are confined in the initial configuration, while the dynamics of
the freezout is not affected by the initial conditions. 
However, to show the impact of the saturation scale we will discuss
both the cases $Q_{0} = 1$ GeV and $Q_0 = 2$ GeV in agreement with the estimates of~\cite{Lappi:2011gu},
for the Pb-Pb collisions simulations at the LHC energy.

\begin{figure}[t!]
\begin{center}
\includegraphics[width=8.5cm]{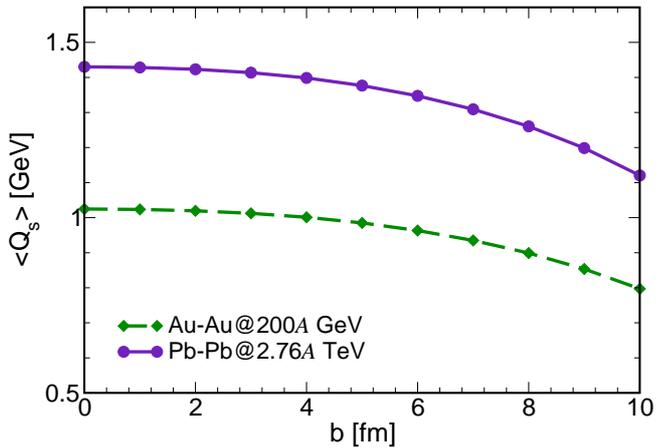}
\caption{\label{Fig:Qsav} (Color online)
Saturation scale $Q_s$ averaged over the transverse plane
for the cases of a Au-Au collision at $\sqrt{s}=200 A$ GeV 
(green dashed line) and a Pb-Pb collision 
at $\sqrt{s}=2.76 A$ TeV (solid indigo line).
All the calculations correspond to
gluons produced at $y=0$ and $p_T=2$ GeV, and $Q_0=1$ GeV.}
\end{center}
\end{figure}

In Fig.~\ref{Fig:Qsav} which we plot the averaged 
value of $Q_s$ on the transverse plane,
\begin{equation}
\langle Q_s\rangle=\frac{\int d^2\bm x_T Q_{s} dN/d{\cal P}}
{\int d^2\bm x_T dN/d{\cal P}}~,
\end{equation}
with $Q_s$ given by Eq.~\eqref{eq:PPPppp} and the weight
function by Eq.~\eqref{eq:densityPS},
for the cases of a Au-Au collision at $\sqrt{s}=200 A$ GeV and a Pb-Pb collision
at $\sqrt{s}=2.76 A$ TeV, for gluons at $y=0$ and $p_T=2$ GeV,
corresponding to $x=0.01$ in the case of the Au-Au collision 
and to $x=7\times 10^{-4}$ in the case of the Pb-Pb collision.

\section{Kinetic theory at fixed $\eta/s$}
In our study we employ transport theory as a base of a simulation code of the fireball expansion created
in relativistic heavy-ion collision~\cite{Ferini:2008he,Plumari:2010ah,Plumari_BARI,Plumari:2012xz}, 
therefore the gluon distribution function $f_1({\bm x}, {\bm p}, t)$
evolves according to the Relativistic Boltzmann Transport (RBT) equation:
\begin{equation}
p_\mu\partial^\mu f_1 = C[f]~, 
\label{eq:BE}
\end{equation}
where $C[f]$ is the collision integral,
\begin{eqnarray}
C[f]&=&\int d\Gamma_2 d\Gamma_{1^\prime} d\Gamma_{2^\prime}
(f_{1^\prime}f_{2^\prime} - f_1 f_2)\nonumber\\ 
&&\times|{\cal M}|^2\delta^4(p_1 + p_2 - p_{1^\prime} - p_{2^\prime})~,
\end{eqnarray}
with $d^3 {\bm p}_{k} = 2 E_{k} (2\pi)^3 d\Gamma_k$, and ${\cal M}$ corresponds to the transition amplitude.

At variance with the standard use of transport theory, in which one fixes
a set of microscopic processes whose scattering matrix is fed into the collision integral, we have developed
an approach that, instead of focusing on specific microscopic calculations or modelings for the scattering matrix, 
fixes the total cross section in order to have the wanted $\eta/s$. 
By means of this procedure we are able to use the Boltzmann equation to simulate
the dynamical evolution of a fluid with specified
shear viscosity, in analogy to what is done within hydrodynamical simulations.
The advantage of the kinetic theory approach at fixed $\eta/s$, compared
to hydro simulations, is twofold: firstly starting from $f(x,p)$, and not from $T^{\mu\nu}$, 
it is direct to incorporate non-equilibrium initial conditions. Secondly we do not
need to specify an ansatz for the deviations $\delta f$ from equilibrium due to
viscous corrections.
This approach to kinetic theory has been also considered in \cite{Huovinen:2008te,El:2009vj}
where it has been shown that transport theory at fixed $\eta/s$ reproduces
the results of viscous hydrodynamics for one-body observable like $T^{\mu\nu}$ or entropy density 
also in the limit in which the system is not in the dilute regime.
This is not surprising because looking at the Boltzmann collision integral in terms of viscosity
allows the analytical derivation of second order viscous hydrodynamics \cite{Grad,Denicol:2012es}.   

In this article we consider only the $2\leftrightarrow 2$ processes 
to compute the collision integral. 
One may think that the introduction of the higher order processes would result naturally
in a change of the available phase space of partons, but once the system is close to an
hydro regime where many collisions happens in a very short time respect to the expansion
time scale and system size, the detail of the single scattering is lost and what matters is
the viscosity of the fluid. Of course one can expect that the microscopic details of the scattering
arises at large $p_T$ where a hydro-like description in terms of a gradient expansion
of the stress tensor breaks down; in the regime
of small and moderate $p_T$, which is the one we are interested to in this study,
the single collision phase space change is not relevant. 
We have indeed checked that we are in a regime where the phase space of the single scatterings 
are not relevant by changing the microscopic two-body scattering matrix from anisotropic to isotropic,
but renormalizing the total cross section in such a way to keep the same $\eta/s$. 
We have found that leaving unchanged all the other parameters,
the elliptic flow is not affected by this change for $p_T \leq 2.5$ GeV, finding some deviations at larger $p_T$ \cite{greco-inpc}. 
A similar  result is seen if $2\leftrightarrow 3$ collisions are switched on and off once again the cross sections
are such to keep $\eta/s$ fixed \cite{greiner-private}.

Once $\eta/s$ is fixed, we compute the total cross section in each cell of the coordinate
space of our grid. To this end we need an analytical relation between $\eta$, temperature,
cross section and density;
as shown in \cite{Plumari:2012ep,Plumari:2012xz}, the Chapmann-Enskog approximation 
supplies such a relation with quite good approximation, in agreement with the results 
obtained using the Green Kubo formula. 
Therefore, we fix $\eta/s$ and compute the pertinent total cross section
by mean of the relation
\begin{equation}
\sigma_{tot}=\frac{1}{15}\frac{\langle p\rangle}{\rho \, g(a)} \frac{1}{\eta/s}=\frac{1}{15} \langle p\rangle \, \tau_{\eta}
~, 
\label{eq:sigma}
\end{equation}
which is valid for a generic differential cross section $d\sigma/dt \sim \alpha_s^2/(t-m_D^2)^2$
as proved in~\cite{Plumari:2012ep}.
In the above equation $a=T/m_D$, with $m_D$ the screening mass regulating the angular dependence
of the cross section, while  
\begin{eqnarray}
g(a)=\frac{1}{50}\! \int\!\! dyy^6
\left[ (y^2{+}\frac{1}{3})K_3(2 y){-}yK_2(2y)\right]\!
h\left(\frac{a^2}{y^2}\right)
\label{g_CE}
\end{eqnarray}
with $K_n$ the Bessel function and $h$ corresponding to the ratio of the transport and the
total cross section $\sigma_{tr}(s)= \sigma_{tot} \, h(m_{D}^2/s)$
and $h(\zeta)=4 \zeta ( 1 + \zeta ) \big[ (2 \zeta + 1) ln(1 + 1/\zeta) - 2 \big ]$. 
The $g(a)$ is the proper function accounting for
the pertinent relaxation time $\tau_{\eta}^{-1}=g(a) \sigma_{tot} \rho$ associated
to the shear transport coefficient. The maximum value of $g$, namely $g( m_D \rightarrow \infty)=2/3$, 
is reached for isotropic cross section and Eq.(\ref{eq:sigma}) reduces to the relaxation time approximation
with $\tau^{-1}_{\eta} =\tau^{-1}_{tr} = \sigma_{tr} \rho$; a smaller value of $g(a)$
means that a higher $\sigma_{tot}$ is needed to reproduce the same value of $\eta/s$. 
However, we notice that in the regime were viscous hydrodynamic applies 
(not too large $\eta/s$ and $p_T$)
the specific microscopic detail of the cross section is irrelevant and our approach 
is an effective way to employ transport theory to simulate a fluid at a given $\eta/s$.
Therefore from one hand we are in hydrodynamical regime from the other that our approach cannot be 
and is not meant as a tool to infer a microscopic description of the QGP. 


\begin{figure}[t!]
\begin{center}
\includegraphics[width=8.5cm]{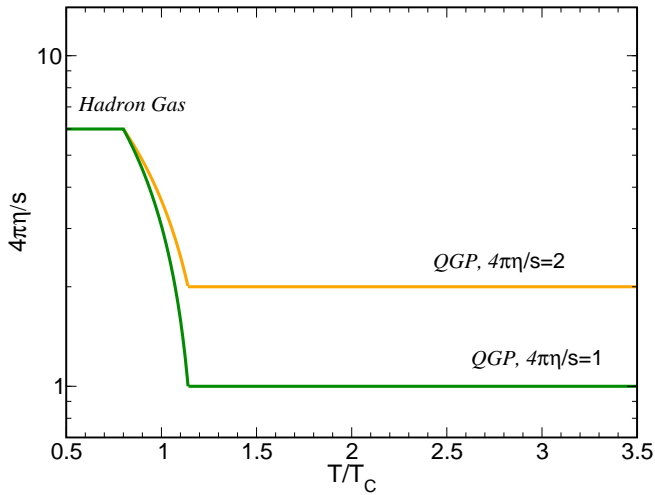}
\caption{\label{Fig:etaSUs} (color online)
Ratio of shear viscosity over entropy density ratio implemented in the present study.
Green solid line corresponds to the case in which we take $4\pi\eta/s=1$ in the quark-gluon-plasma
phase, which we then connect smoothly to the value in the hadron phase by means
of a kinetic freezout.
For completeness we also show the orange line corresponding to the case
$4\pi\eta/s=2$ in the quark-gluon-plasma phase, which is once again connected smoothly
to the value in the hadron phase.}
\end{center}
\end{figure}
The shear viscosity over entropy density ratio which we use in our study is plot in Fig.~\ref{Fig:etaSUs}.
In the plasma phase $\eta/s$ is taken to be a constant, whose numerical value is fixed case by case:
in the figure the cases $4\pi\eta/s=1$ and $4\pi\eta/s=2$ are represented, 
but we also perform some simulations with larger values of the ratio.
We implement a kinetic freezout by assuming that $\eta/s$ increases smoothly in a temperature range from the
plasma phase to a hadron phase values which is fixed by referring to the estimates 
in~\cite{Prakash:1993bt,Chen:2007xe,Demir:2008tr}. 
In this way we take into account scatterings in the hadron phase as well, which however give a very tiny
contribution to the collective flow because of the damping due to the larger viscosity and the reduce
eccentricity at this later stage.

\begin{figure}[t!]
\begin{center}
\includegraphics[width=8.5cm]{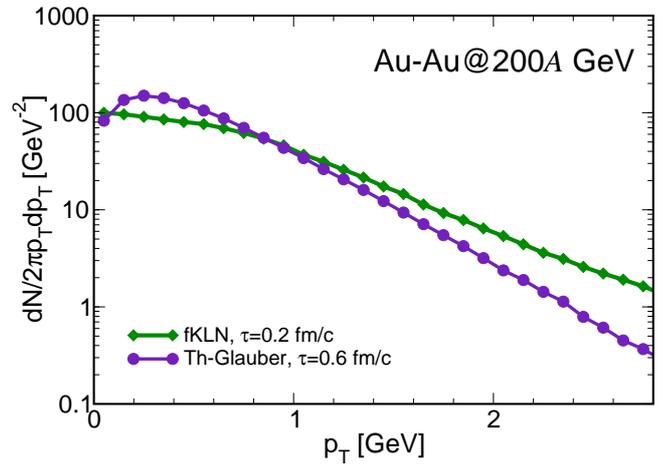}
\caption{\label{Fig:spettri_CGC} (Color online)
Initial $p_T-$spectra at midrapidity for the case of a Au-Au collision
at $\sqrt{s} = 200$ GeV, with an impact parameter $b=7.5$ fm.}
\end{center}
\end{figure}

In the following, we will consider three different types of initial distribution function in the phase-space,
which have been considered also in our previous study~\cite{Ruggieri:2013bda}:
two of them are commonly employed in hydro simulations, while the third one
represents the novelty of the present study and relies on the ability
of the transport approach to include the saturation scale in the initial 
distribution function, as shown in Fig.~\ref{Fig:spettri_CGC}. For simulations at the RHIC energy 
we refer to Au-Au collision at $\sqrt{s}= 200 \, A$GeV; we present here results
for $b=5.2$ fm, $b=7.5$ fm and $b=9$ fm. 
For simulations at the LHC energy we will refer to Pb-Pb collisions at $\sqrt{s}= 2.76 \, A$TeV,
focusing on the case $b=7.5$ fm.

The standard initial condition for simulations of the plasma fireball created at RHIC
is based on the Glauber model, with an $\bm x$-space distribution given by the a standard mixture $0.85N_{part}+0.15N_{coll}$
and a $\bm p$-space thermalized spectrum in the transverse plane at a time $\tau_0 = 0.6\rm \, fm/c$ with
a maximum initial temperature $\rm T_0 = 340 \rm MeV$. 
Following the nomenclature introduced in~\cite{Ruggieri:2013bda} we will refer to this case as Th-Glauber 
corresponding to full circles in Fig.\ref{Fig:spettri_CGC}.
Within previous studies based on hydro simulations 
the impact of an initial factorized glasma state has been performed considering an $\bm x$-space
distribution given by the fKLN (or MC-KLN), while in the momentum space 
the spectrum has been considered thermalized at $\tau_0 \sim 0.6 \div 0.9$ fm/c~\cite{Luzum:2008cw,Song:2011hk}; 
we refer to this case as Th-fKLN; among other things this initial condition, 
when implemented in hydro simulations,
leads to the conclusion that the liquid created by the melting of the glasma is characterized
by $4\pi \eta/s \sim 2$~\cite{Luzum:2008cw,Song:2011hk,Adare:2011tg} at RHIC,
at least for non-central collisions. 
The third initial conditions is the full fKLN initial conditions where,
beyond the $\bm x$- space, the saturated distribution in $\bm p$-space
proper of the model is implemented as well, see Fig.~\ref{Fig:spettri_CGC} solid thick line with diamonds.
We have verified that our main results on thermalization times and elliptic flows are not 
affected by the choice of $\tau_0$. 
In this case as initial time we take $\tau_0=0.2$ fm/c with no assumption  
about thermalization in the transverse plane, as it should the case of any implementation
of the KLN model since its proper momentum distribution is out of equilibrium. 
This initial condition is not usually considered in hydrodynamics because there it is implicitly assumed 
the system is locally thermalized in the transverse plane, and in fact the initial transverse
energy density profile is connected to the initial $\bm x$-space fKLN distribution 
assuming equilibrium thermodynamic relations.  In the context of viscous hydrodynamics 
it is possible to include initial non-equilibrium
conditions by introducing an initial non-vanishing value
for the viscous tensor $\Pi_{\mu\nu}(\tau_0)$. 
This has been mostly studied for the shear stress tensor and it has been seen to
have a quite small impact on the $v_2$ at least on the bulk of the system~ \cite{Luzum:2008cw,Song:2007fn,Niemi:2011ix}. 
However the non-equilibrium implied by fKLN implies a change in the trace of the energy-momentum
tensor that could be related in hydrodynamical language to a finite bulk stress tensor at $\tau_0$.
To our knowledge it has never been investigated the relation between an initial
non vanishing bulk $\Pi_{\mu\nu}(\tau_0)$ and the initial non-equilibrium implied 
by the saturation scale in the distribution function $f(\bm x, \bm p)$. 

Similarly we introduce Th-Glauber, Th-fKLN and fKLN initial conditions for the
LHC runs. In the case of Th-Glauber and Th-fKLN the maximum temperature is taken
to be $T_0 = 550$ MeV and the system is assumed to be thermalized in the transverse
plane at $\tau_0=0.3$ fm/c; for the case of the fKLN initial condition, we keep
$\tau_0=0.2$ fm/c. This time should correspond to the time interval in which 
the color strings decay forming the strongly interacting parton liquid.

\begin{figure}[t!]
\begin{center}
\includegraphics[width=8.5cm]{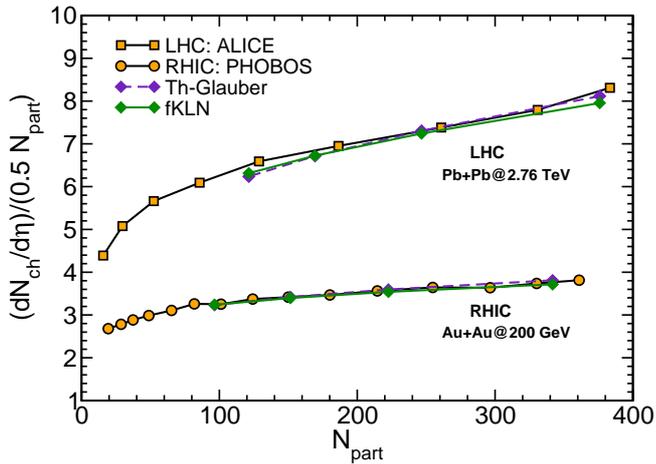}
\caption{\label{Fig:multe}
(Color Online) Multiplicity as a function of $N_{part}$ at RHIC and LHC energies as indicated in the figure.
Experimental data are taken from Refs.~\cite{PHOBOS,ALICE}.}
\end{center}
\end{figure}


The particle multiplicity is fixed to correctly reproduce the experimental 
one for all the three cases. 
This is shown in Fig.~\ref{Fig:multe} where we plot the particle multiplicity, measured at central rapidity
in units of the number of pairs, as a function of 
the number of participants, for the cases of the fKLN and Th-Glauber initializations. 
Experimental data are taken from Refs. \cite{PHOBOS,ALICE}.
In the case of the Th-Glauber and Th-fKLN we have assumed boost invariance in the
longitudinal direction for the initial configuration; 
on the other hand, for the case of the fKLN initial condition
a small $y-$dependence comes from the distribution in Eq.~\eqref{eq:densityCS}.
Nevertheless in both the calculations the total number of particles
is the same, as it is shown in Fig.~\ref{Fig:multe}.
\begin{figure}[t!]
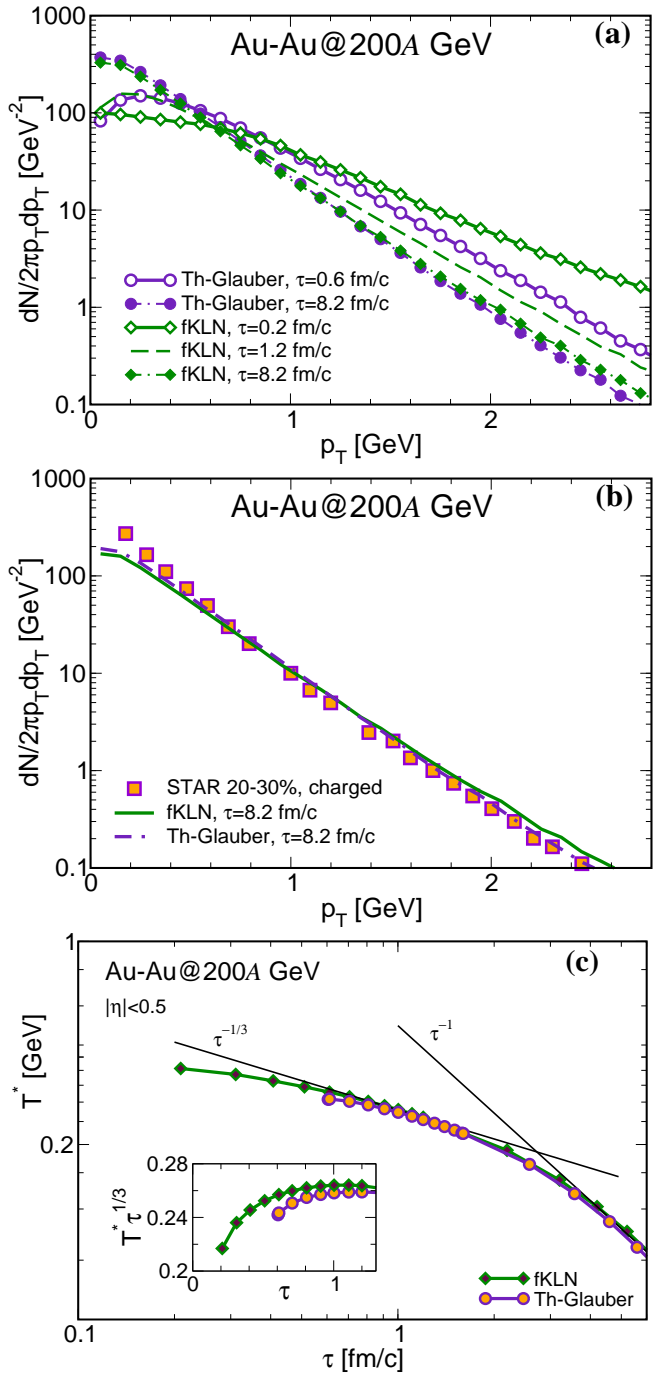

\begin{center}
\includegraphics[width=8.5cm]{figures/spe_evol.eps}\\
\includegraphics[width=8.5cm]{figures/spe_RHIC_data.eps}\\
\includegraphics[width=8.4cm]{figures/temp_evol.eps}
\caption{\label{Fig:spettri_CGC2} (Color online)
{\em Upper panel:} Time evolution of spectra.
{\em Middle panel:} Comparison of the final spectra obtained by simulations
with experimental data 
\cite{STAR-spe}. 
{\em Lower panel:} Time evolution of temperature $T=E/3N$. 
In the inset we plot $T\tau^{1/3}$ for the early
stage of the collision. 
In all panels $b=7.5$ fm.
The Th-fKLN case is not shown because for both spectra and temperature
evolution we do not find visible deviations from the Th-Glauber 
initialization.}
\end{center}
\end{figure}

\section{Thermalization and Isotropization}
In Fig.~\ref{Fig:spettri_CGC2} we collect the initial spectra, $dN/2\pi p_T d p_T$,  
integrated over the momentum rapidity window $|y|<0.5$,
for the case of a Au-Au collision $\sqrt{s}=200 A$ GeV  
for the fKLN and Th-Glauber initial conditions at their respective initial
times $\tau_0$ and at the final time $\tau=8.2$ fm/c.
For the case of the Th-fKLN, we find that the spectra are the same as in the case
of the Th-Glauber, therefore we do not plot them in the figure. 
We have also shown by the dashed line the spectrum at $\tau=1.2$ fm/c for the case
of the fKLN initial condition. 

Our main aim is to focus on the impact of the initial conditions implied
by the fKLN on the elliptic flow; however we discuss before the time evolution
of spectra and the related isotropization.
The results in Fig.~\ref{Fig:spettri_CGC2} are obtained with $4\pi\eta/s=1$.
We notice that initially the fKLN spectrum is quite far from a thermalized spectrum; in fact,
it embeds the saturation effects which should be proper of the melted glasma.
Nevertheless the spectrum thermalizes in the transverse plane within $1$ fm/c,
since its $p_T$ dependence becomes exponential with a slope very similar to the Th-Glauber.

In the middle panel we compare our results with the experimental data
of the STAR collaboration for the charged hadrons \cite{STAR-spe}. 
In order to 
make this comparison we have assumed the quark-hadron duality
which implies that for each parton a single hadron is produced; 
moreover we have multiplied our results by the factor
$2/3 \times 3/4$ which takes into account the facts that experimental data
correspond to charged hadrons only, and that the rapidity range of experimental 
data is different from ours.
We want only to show that the final spectra between Th-Glauber and fKLN have a very simlar
shape which make more meaningful the comparison of the elliptic flow discussed in the next Section.
The data are mainly shown to have a reference. 

In the lower panel of Fig.~\ref{Fig:spettri_CGC2} (c), we have plot the ratio $T^*=E/3N$,
evaluated in the local rest frame of the fluid and
representing the temperature in the case of a thermalized system, as a function of time.
The change of slopes of the curves suggest that initially the cooling of the fireball
is dominated by the longitudinal expansion; as soon as the transverse expansion begins 
($\tau \sim 2-3\, \rm fm/c$), the cooling becomes faster and indeed the fireball reaches the freezout energy density 
within few fm/c.
In the inset of Fig.~\ref{Fig:spettri_CGC2}, 
we plot the quantity $T^* \cdot \tau^{1/3}$.
In the case of 1D expansion without dissipation, 
implies that a thermalized system $T\propto\tau^{-1/3}$. 
We find that in the case of the fKLN (solid green line) 
the product $T^* \cdot \tau^{1/3}$ is strongly dependent
on time because the system initially is quite far from equilibrium,
and the initial non-equilibrium generates some entropy to thermalize; 
however at $\tau \sim 0.8 \rm \, fm/c$ we find  $T^* \cdot \tau^{1/3}$  saturates to a 
constant value, signaling the system has thermalized in the whole momentum space. 
We notice a tiny evolution 
also for these cases that we have indicated as thermal. The reason is that the initial spectra are
thermal only in the transverse plane, 
while they are boost-invariant along the longitudinal direction.
Furthermore as one can expect my solving the Boltzmann under boost invariance 
at mid-rapidity $T^\star$ should decrease as $\tau^{-\delta}$ with $\delta= P_L/\epsilon$
that is smaller than $1/3$ \cite{Huang:2014iwa}. 
This causes a tiny evolution to the constant value,
that would disappear when the distribution is thermal also in the longitudinal direction.
At larger times $T^\star$ decreases faster because the 3D expansion sets in
and  at $\tau \sim 2-3 \, fm/c$ one has $T^\star\sim\tau^{-1}$ as shown in the main panel
of Fig.~\ref{Fig:spettri_CGC2} with the thin solid line indicating the $\tau^{-1}$ behavior. 
We remark that even if we have chosen to start
our simulation at $\tau_0 = 0.2$ fm/c, our final result on thermalization
time required by the KLN initial condition is unaffected if we shift the
initial time to a reasonably larger value.

A posteriori our results on short thermalization times are quite natural:
in fact, we have assumed the fluid expands with a very small shear viscosity,
which naturally implies that the coupling among the partons are nonperturbative,
hence leading efficiently the distribution to the fixed point of the Boltzmann equation,
namely a thermalized distribution.
These results on fast thermalization are in agreement with 
earlier studies showing that two-body collisions are insufficient
to achieve a fast thermal equilibrium \cite{El:2007vg}, and three-body processes
are necessary. 
In fact in that case the perturbative QCD two-body cross section 
is used, which corresponds to $\eta/s$ about one order of magnitude larger than in our case;
hence, to achieve fast thermalization three-body processes have to be fed into the
collision integral. 
The difference with \cite{El:2007vg} is that we do not focus on the microscopic
details which lead to thermalization, as discussed in Section III: we are interested to simulate
the dynamics of a fluid with few macroscopic properties specified, 
namely the equation of state and the ratio $\eta/s$. Thus we normalize the
cross section locally to get the wanted $\eta/s$, and our scattering rates 
even if obtained effectively by two-body scattering are very large, 
leading eventually to a fast thermalization.

\begin{figure}[t!]
\begin{center}
\includegraphics[width=8.3cm]{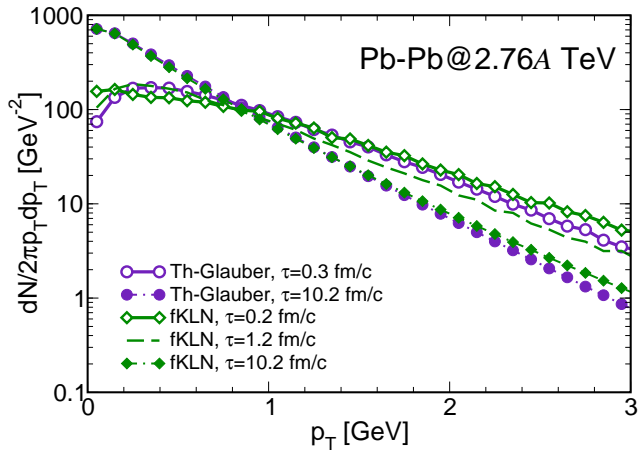}\\
\caption{\label{Fig:spettri_CGC3} 
{\em (Color online)} Time evolution of spectra at midrapidity for Pb-Pb collision
at $\sqrt{s} = 2.76 A$ TeV, with an impact parameter $b=7.5$ fm.
}
\end{center}
\end{figure}
In Fig.~\ref{Fig:spettri_CGC3}, we plot our results for the spectra for a Pb-Pb collision
at $\sqrt{s}=2.76 A$ TeV, corresponding to a typical LHC collision. As expected, the qualitative behaviour
of spectra is unchanged shifting from the RHIC to the LHC energy.
In this case however the effect of the saturation scale is smaller, because the
initial temperature for the Th-Glauber initial condition is larger, as it can be inferred
from the spectra. Hence the $p_T-$spectra for the thermalized case is indeed quite close to the fKLN one,
as shown in Fig.\ref{Fig:spettri_CGC3} (upper panel).
Also in the case of the Pb-Pb collision, starting from $\tau_0=0.2$ fm/c
for the fKLN melted glasma initial condition, time evolution of spectra shows that the transverse
thermalization occurs within $\Delta\tau \approx 1$ fm/c.
At LHC the time scale at which $T^\star \cdot \tau^{1/3}$ is constant appears to be quite similar to the one at RHIC
in Fig.\ref{Fig:spettri_CGC2} (c).

\begin{figure}[t!]
\begin{center}
\includegraphics[width=8.3cm]{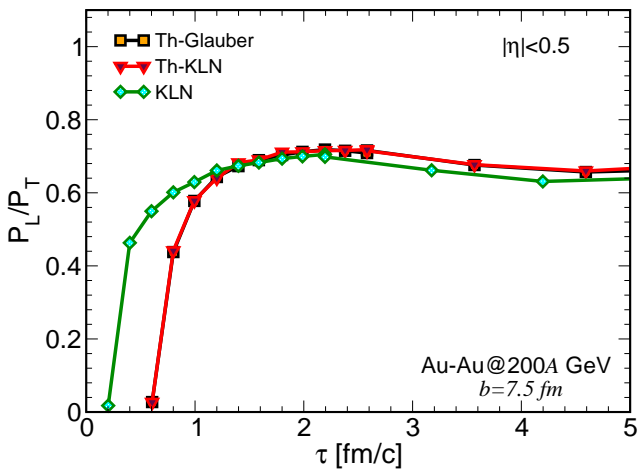}\\ 
\caption{(Color online) Time evolution of the ratio $P_L/P_T$, for several initial conditions.
Upper panel results refer to Au-Au collisions
at $\sqrt{s} = 200 A$ GeV. 
}
\label{Fig:Ptot2}
\end{center}
\end{figure}

We have studied also the isotropization of pressures for the expanding system. 
The energy-momentum tensor is defined locally as
\begin{equation}
T^{\mu\nu}(x)=\int\frac{d^3 \bm p}{(2\pi)^3}\frac{p^\mu p^\nu}{E}
f(x,p)~,
\label{eq:Tmunu1}
\end{equation}
where $f$ corresponds to the invariant distribution function. 
In our simulations the energy-momentum tensor is defined
in each cell and it is computed
in the local rest frame of the fluid, subtracting the radial flow contribution; 
we then define locally transverse and longitudinal pressures, $P_T$ and $P_L$ respectively, as
\begin{eqnarray}
P_T(x)&=&\frac{T_{xx}(x) + T_{yy}(x)}{2}~,\\ 
P_L(x)&=&T_{zz}(x)~;
\end{eqnarray}
finally we average $P_L/P_T$ over several cells in order to lower 
the effect of statistical fluctuations.
In Fig.~\ref{Fig:Ptot2} we plot our results about $P_L/P_T$ as a function
of time, for several initial conditions and for the case of strong coupling
$4\pi\eta/s=1$. In particular in the figure we have averaged
over a square in the central region of the transverse plane whose side is of $2$ fm,
while in the longitudinal direction we have averaged over $|\eta|<0.5$. 
At early times the longitudinal pressure is zero
due to the Bjorken initial conditions.
Our findings suggest that independently on the initial condition
implemented, the system becomes nearly isotropic ($P_L/P_T \sim 0.7$)
within a $\Delta\tau_{isotr} \sim 0.5\,\rm fm/c$
for $4\pi\eta/s=1$. 

\begin{figure}[t!]
\begin{center}
\includegraphics[width=8.3cm]{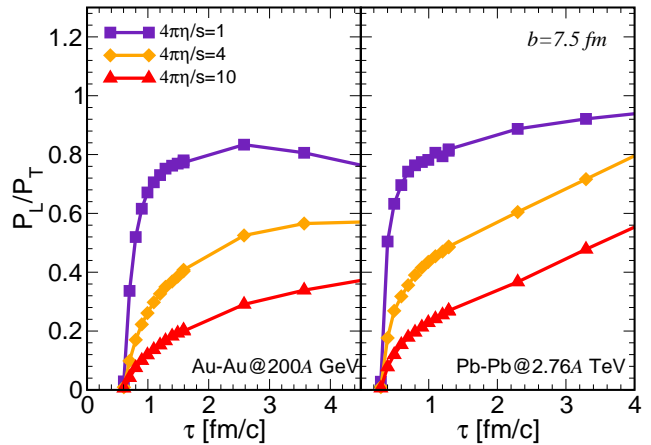}
\caption{(Color online) Time evolution of the ratio $P_L/P_T$, for the case of the Th-Glauber initial condition,
for several values of the ratio $\eta/s$. Left panel corresponds to Pb-Pb collision 
at $\sqrt{s}=2.76 A$ TeV; right panel corresponds to Au-Au collision at $\sqrt{s}=200 A$ GeV.
Selection in transverse plane is $|R_T|<2 \rm \,fm$ and space-time rapidity $|\eta|<0.5$.
All the calculations refer to collisions with an impact parameter $b=7.5$ fm.}
\label{Fig:Ptot3}
\end{center}
\end{figure}

We have selected the region $|\eta|<0.5$ checking that the particle in this space region are nearly
correspondent to all the particles 
with momentum rapidity $|y|<1$ to which we are mainly interested.
We have also checked that the maximum value of $P_L/P_T$ marginally depends on the space region 
selected while its maximum values increase up to $P_L/P_T \sim 0.8$ if one selects the central 
region (see also Fig.\ref{Fig:Ptot3}), while it goes down to $P_L/P_T \sim 0.6$ if one considers the entire fireball in both transverse
and longitudinal direction.
A similar behavior is observed at LHC as can be seen for the Th-Glauber case also in the left panel
of Fig.\ref{Fig:Ptot3}.

It is interesting to compute how the isotropization is affected
by the $\eta/s$ of the fluid. To this end
in Fig.~\ref{Fig:Ptot3} we show our results for pressures
at the RHIC (right panel) and LHC (left panel) energies, for the Glauber
initial condition (for the fKLN initial condition we obtain similar results),
for different values of the ratio $4\pi\eta/s$. In particular,
the value $4\pi\eta/s =10$ corresponds approximately to the perturbative
QCD estimate~\cite{Arnold:2003zc,Plumari:2012xz,Plumari:2012ep}. As expected, the larger the 
$\eta/s$, the smaller the ability of the system to remove the initial 
pressure anisotropy and a perturbative fluid would not be able
to sufficiently isotropize the pressure. The ratio $P_L/P_T$ reaches
its maximum value slowly as $\eta/s$ increases: turning from $4\pi\eta/s=1$
to $4\pi\eta/s=10$ 
the capability of the system to reach isotropization is lost while it is barely
reached for $4\pi\eta/s \sim 0.3$ but $\Delta\tau_{isotr}\approx 3\,\rm fm/c $.
At LHC the trend is similar but the larger density to reach slightly larger values
of $P_L/P_T$.

Similar results on incomplete isotropization have been found in~\cite{Ryblewski:2013eja},
where Schwinger mechanism causes the color strings decaying into a parton liquid and 
shear viscosity is taken into account via the relaxation time ansatz.
Respect to~\cite{Ryblewski:2013eja} we consider a full 3D+1 simulation
and we do not employ the relaxation time approximation, solving
the kinetic equation with the full Boltzmann integral. On the other hand,
we miss the field dynamics that causes oscillations of $P_L/P_T$ which 
in the early phase generates a peak in the $P_L/P_T$ at $\tau \sim 1\,\rm fm/c$
and in 3D+1 could also
lead to instabilities in three dimensions in the very early stage of the expansion. 

The results in Fig.~\ref{Fig:Ptot2} and \ref{Fig:Ptot3} are comparable with previous 
calculations~\cite{Gelis:2013rba,Ryblewski:2013eja} in which a dynamics for
the color glasma fields is introduced; this comparison is quite meaningful
because it is well known that in the very initial stage of its dynamics,
the glasma is characterized by $P_L<0$ due to strong fields in the
longitudinal direction. This negative pressure cannot be reproduced within
our simulation code at the moment because we have not yet implemented a dynamics for
the classical chromoelectric and chromomagnetic fields.
Nevertheless as shown in~\cite{Gelis:2013rba,Ryblewski:2013eja}
the ratio $P_L/P_T$ becomes positive within $\tau_+\approx 0.1\div 0.2$ fm/c
depending on the coupling strength, and unless the system is in weak coupling
the strength of the fields becomes negligible for $\tau > \tau_+$~\cite{Ryblewski:2013eja},
which justifies our assumption that starting the simulations at $\tau=\tau_0 = 0.2$ fm/c
the evolving fireball can be described by a parton liquid with positive pressure.

\section{Elliptic flow}

The main goal of our study is the computation of the differential elliptic flow
for different initial conditions, both at RHIC and LHC energies. 
The main motivation has been discussed in the first part of Section III.

\begin{figure}[t!]
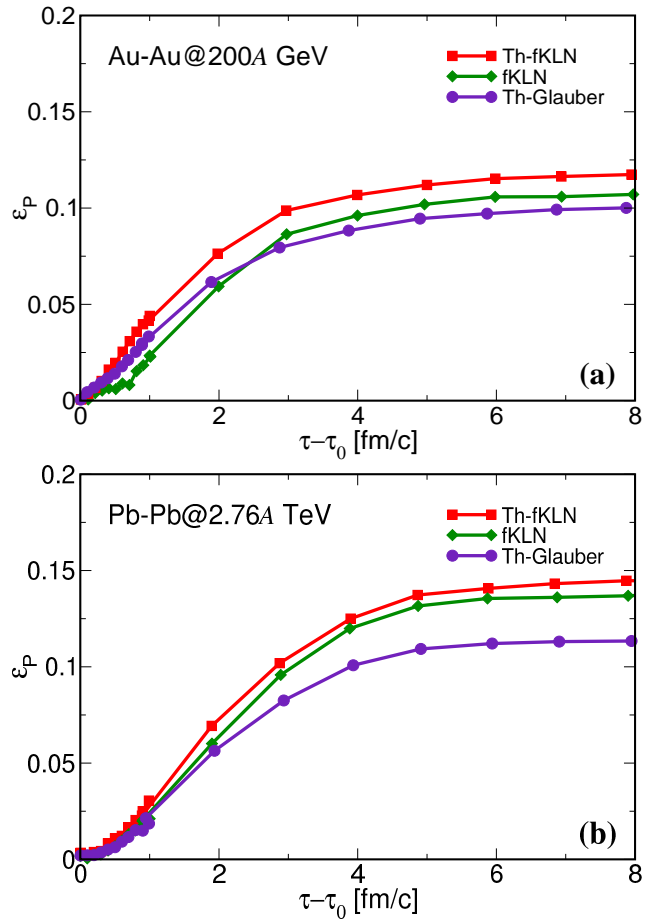

\begin{center}
\includegraphics[width=8.3cm]{figures/epsP_RHIC.eps}\\
\includegraphics[width=8.3cm]{figures/epsP_LHC.eps}
\caption{(Color online) Momentum eccentricity, $\varepsilon_P$, as a function of time
for three different initial conditions. 
Upper panel refers to Au-Au collisions
at $\sqrt{s} = 200 A$ GeV; lower panel to Pb-Pb collisions
at $\sqrt{s} = 2.76 A$ TeV. In all calculations the impact parameter $b=7.5$ fm,
$4\pi\eta/s=1$. Components of the energy-momentum tensor have been
integrated over $|\eta|<0.5$. For the fKLN initial condition we have chosen $\tau_0=0.2$ fm/c,
while for the thermal distributions $\tau_0=0.6$ fm/c.}
\label{Fig:eP}
\end{center}
\end{figure}
\begin{figure}[t!]
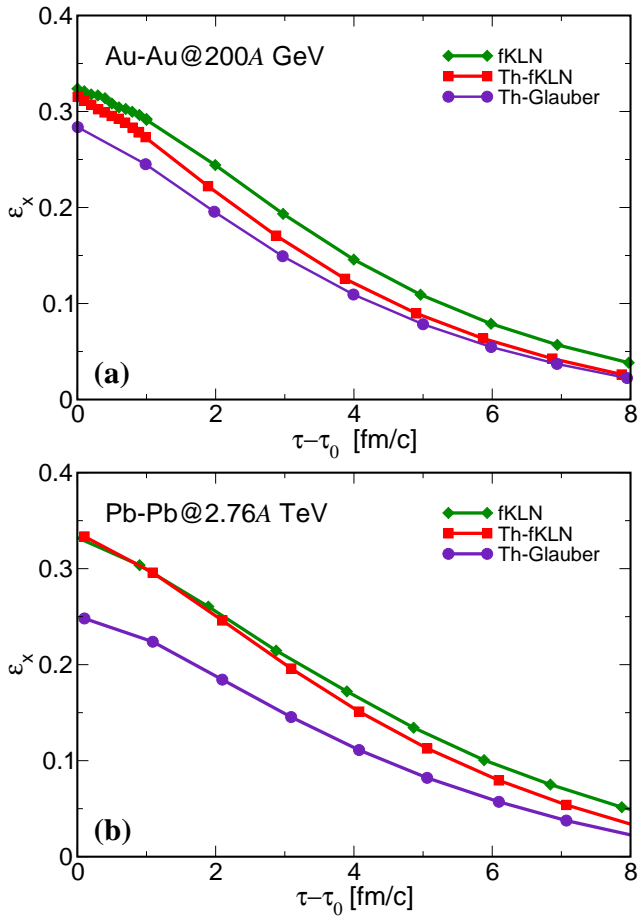

\begin{center}
\includegraphics[width=8.3cm]{figures/ecceR.eps}\\
\includegraphics[width=8.3cm]{figures/ecce.eps}
\caption{Time evolution of eccentricity, for the cases of a Au-Au collision (upper panel)
at $\sqrt{s}=200 A$ GeV, and of a Pb-Pb collisions
at $\sqrt{s} = 2.76$ TeV. In all calculations the impact parameter $b=7.5$ fm
and $4\pi\eta/s=1$. For the fKLN initial condition we have chosen $\tau_0=0.2$ fm/c,
while for the thermal distributions $\tau_0=0.6$ fm/c.}
\label{Fig:ecce}
\end{center}
\end{figure}

Before discussing the elliptic flow we show the results obtained for the
momentum eccentricity,
\begin{equation}
\varepsilon_P = \frac{\langle T_{xx}\rangle - \langle T_{yy}\rangle}
{\langle T_{yy}\rangle + \langle T_{xx}\rangle}~,
\label{Eq:ePdef}
\end{equation}
which is usually computed also in hydrodynamics~\cite{Romatschke:2009im,Teaney:2009qa,Shen:2011kn,Song:2008si}
and whose evolution is not related to
the freezout by the Cooper-Frye hypersurface 
as insted happens for $v_2$. 
The average in the above equation is understood in the transverse plane at midrapidity.
We compute this quantity to show that especially for the bulk of the system
the kinetic Boltzmann approach determines a hydro-like evolution.
In Fig.~\ref{Fig:eP}, we plot the momentum eccentricity
as a function of time, for the three different initial conditions we study in this article,
for both RHIC and LHC energies. For completeness in Fig.~\ref{Fig:ecce} we plot
the space eccentricity $\varepsilon_x$ as a function of time for the cases
of collisions at RHIC (upper panel) and LHC (lower panel) energies.
Results in Fig.~\ref{Fig:eP} show that momentum anisotropy is built-up within the first few fm/c,
which means that the elliptic flow is being developed in the same
time range. However, in this time range $\varepsilon_x$ in the case
of the fKLN initial condition is comparable to that of the Th-fKLN.
Eventually the system enters the freezout and almost free streaming regime, 
and the $\varepsilon_x$ for the two cases evolve in a different way.
This suggests that when the major part of the momentum anisotropy
is formed, the fireball expands in the transverse plane almost in the same
way both in the Th-fKLN and in the fKLN cases; therefore the difference in elliptic flow
in the two cases has to be generated mainly by the
different momentum distribution in the initial stage. 

Our results on $\varepsilon_P$ are in fair agreement with 
those obtained within hydro simulations~\cite{Shen:2011kn,Song:2008si}.
A strict quantitative comparison with the aforementioned results 
is not feasible because the results in~\cite{Shen:2011kn,Song:2008si} 
are obtained using simulations in 2+1 dimensions with some differences in the initial 
temperature and density profile and a different temperature dependence
of the hadron gas shear viscosity and also a different freezout 
temperature. Even with these differences 
we can compare our results with those of the EOS I case (perfect gas EoS) in~\cite{Song:2008si},
see their Fig.~2 .
The time evolution of $\varepsilon_P$ has some difference in the two cases,
in particular in our case it grows faster with time,
which is likely due to the differences mentioned above,
nevertheless the asymptotic value are in good agreement with the Th-Glauber case.

\begin{figure}[t!]
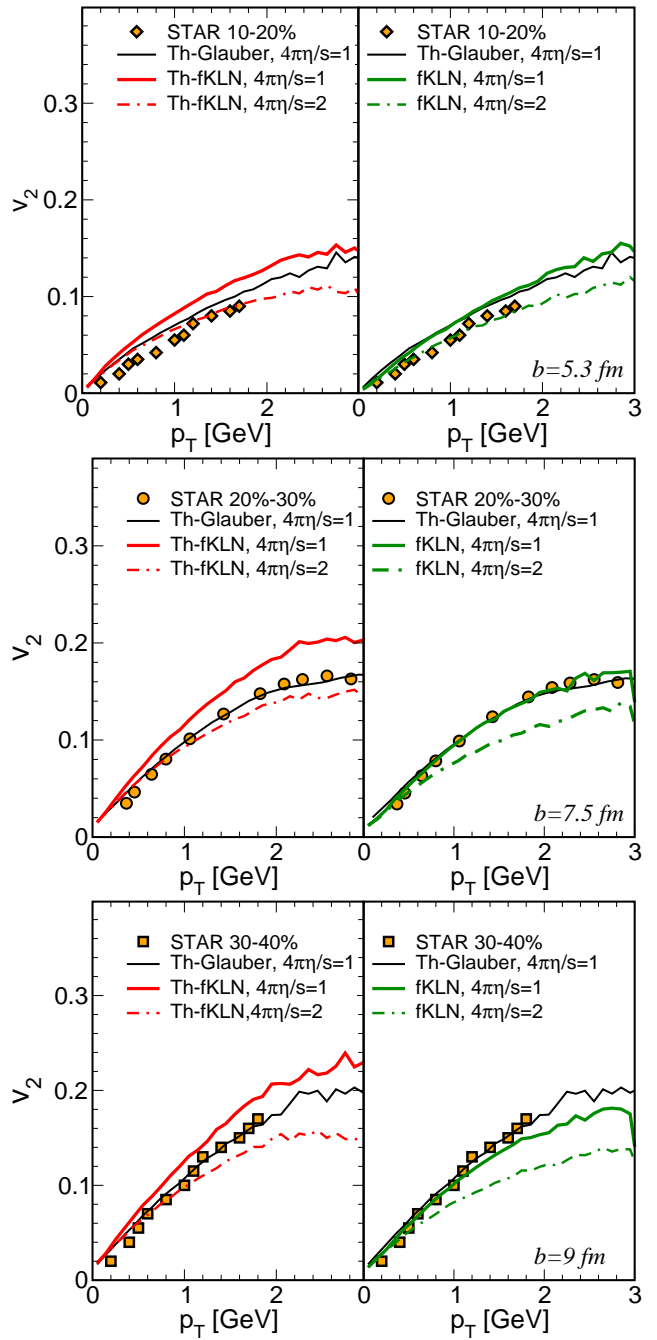

\begin{center}
\includegraphics[width=8.3cm]{figures/v2_pt_fKLN_2.eps}\\
\includegraphics[width=8.3cm]{figures/v2_pt_fKLN_1.eps}\\
\includegraphics[width=8.3cm]{figures/v2_pt_fKLN_3.eps}
\caption{(Color online) Elliptic flow $v_2(p_T)$ at midrapidity $|y|<0.5$ for different initial conditions 
and $\eta/s$ as in the legend.
All the calculations refer to Au-Au collisions
at $\sqrt{s} = 200$ GeV. From the upper to the lower panel the impact parameter 
$b=5.3$ fm, $b=7.5$ fm and $b=9$ fm respectively.}
\label{Fig:v2pt}
\end{center}
\end{figure}

In Fig.~\ref{Fig:v2pt}, we collect our results for the differential 
elliptic flow for the case of a Au-Au collision at $\sqrt{s}=200 A$ GeV,
with three different impact parameters: from the upper to the lower panel the impact parameter is
$b=5.3$ fm, $b=7.5$ fm and $b=9$ fm respectively. The case $b=7.5$ fm 
corresponding to the $20-30\%$ centrality class at RHIC has been 
already considered in~\cite{Ruggieri:2013bda}, but with a different value of $Q_0$:
here  $\langle Q_0\rangle\approx 1$ GeV in the transverse plane at $x=0.01$,
while $\langle Q_0\rangle\approx 1.4$ GeV in~\cite{Ruggieri:2013bda}.
We have used a different value of $Q_0$ in the present calculations mainly because
in this way we do not need to manipulate other parameters, among them the initial temperature
as well as the T dependence of $\eta/s$ in the cross-over region that regulates the kinetic freezeout temperature, 
in order to obtain final spectra which are in agreement among the different initial conditions; moreover we want to check 
that within a reasonable uncertainty on $\langle Q_s\rangle$
the effect we measure on the elliptic flow persists.

Spectra corresponding to these data are plot in Fig.~\ref{Fig:spettri_CGC2}
for the case $b=7.5$ fm (for the other values of $b$ we obtain similar results).
In Fig.~\ref{Fig:v2pt} we have split the results for each value of $b$ 
into two panels: the left one corresponds to our
hydro-like calculations, in which the initial distribution is assumed to be
thermalized in the transverse plane. On the right panel we plot the results
obtained assuming the nonequilibrium distribution in the initial condition. 
To guide the eye, in the figure we also plot experimental data for $v_2$ in
the relevant centrality class from the STAR collaboration~\cite{Adams:2004bi}.
The $v_2(p_T)$ from the kinetic simulations are quite close to the data for the
Th-Glauber and fKLN initializations with $4\pi\eta/s=1$. However we remind that
no hadronization process is yet included in our approach, so the agreement with the charged
hadrons has to be taken with caution even if it indicates that the azimuthal 
asymmetries generated are in the right ball park.

The Glauber initial condition is in agreement with experimental data for $4\pi\eta/s=1$; 
instead, in the case of Th-fKLN the elliptic flow for the same value
of $\eta/s$ overshoots data except for very central collisions: in order to reproduce 
better experimental data one, $\eta/s$ has to be increased by about a factor of two.  
These results are in agreement with the ones obtained from viscous hydrodynamics 
\cite{Song:2011hk,Adare:2011tg,Luzum:2008cw}, showing the solidity and
consistency of our transport approach at fixed $\eta/s$. 
The necessity of a larger $\eta/s$ to reproduce data on elliptic flow is
usually understood in terms of the larger initial eccentricity of the
Th-fKLN fireball compared to the Glauber one, which would result in fact in 
a larger anisotropy in momentum space unless viscosity is large enough
to damp the flow~\cite{Drescher:2006pi,Luzum:2008cw,Song:2011hk}.
 
\begin{figure}[t!]
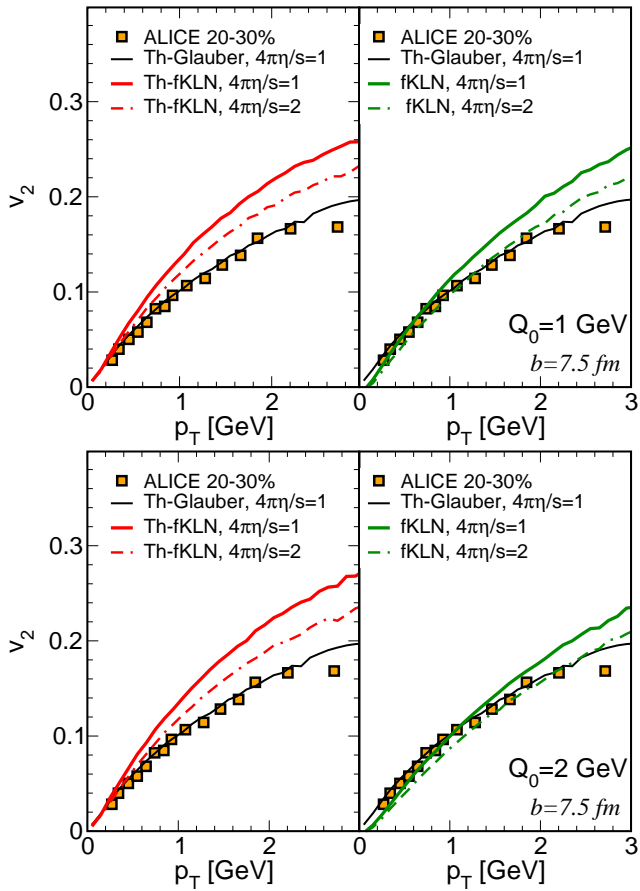

\begin{center}
\includegraphics[width=8.3cm]{figures/v2_pt_LHC.eps}\\
\includegraphics[width=8.3cm]{figures/v2_pt_LHC_2.eps}
\caption{(Color online) {\em Upper panel.} Elliptic flow $v_2(p_T)$ for the different initial conditions and $\eta/s$ as in the legend,
for the case $Q_0=1$ GeV where $Q_0$ is defined in Eq.~\eqref{eq:PPPppp}.
{\em Lower panel.} Same quantity for $Q_0=2$ GeV.
All the calculations refer to Pb-Pb collisions
at $\sqrt{s} = 2.76$ TeV, with an impact parameter $b=7.5$ fm.}
\label{Fig:v2ptL}
\end{center}
\end{figure}

In the right panel of Fig.~\ref{Fig:v2pt}, we present the 
result for the fKLN model, when the proper distribution function is implemented 
in both the $\bm x$ and $\bm p$ spaces. 
We find that fKLN with a $4\pi\eta/s=1$ gives a $v_2(p_T)$ quite similar
to the Th-Glauber, even if the initial eccentricity in this case
is the same as the one of the Th-fKLN case. 
For fKLN with $4\pi\eta/s=2$ the differential elliptic flow would be too small.
In other words the initial out-of-equilibrium fKLN distribution reduces the efficiency
in converting $\epsilon_x$ into $v_2$. 
Our interpretation is that the initial large eccentricity of the fKLN configuration
is compensated by the key feature of an almost saturated initial distribution 
in $\bm p$-space below the saturation scale $Q_s$ and probably by the softer tail at 
$p_T>Q_s$. We notice that changing the centrality class we have found 
that the effect of the initial distribution on the elliptic flow 
becomes smaller as we turn from peripheral to central collisions, as it can be seen
in Fig.~\ref{Fig:v2pt}. 
We obtain similar results at the LHC energy, Fig.~\ref{Fig:v2ptL}, where we collect the results
of our computation of $v_2(p_T)$ for $b=7.5$ fm and for $Q_0=1$ GeV (upper panel)
and $Q_0=2$ GeV (lower panel), in order that $\langle Q_s\rangle$ is in agreement 
with the estimates of~\cite{Lappi:2011gu}.
We can see that the Th-fKLN overestimate the $v_2(p_T)$ at both $4\pi\eta/s=1$ 
and $2$ and probably
even a $4\pi\eta/s \sim 3$ is needed to account for the data.
When the full fKLN is implemented we can see that the damping effect already discussed at RHIC
is such that with an $4\pi\eta/s=2$ one predicted a $v_2(p_T)$ quite close to experiments.

This result we measure on the elliptic flow can be understood naively connecting
$v_2$ to the slope of the spectrum, which corresponds to the inverse temperature
if the spectrum is a thermal one.
In fact, the elliptic flow can be understood as a larger slope of the momentum spectrum in the 
out of plane $\vec x$ direction respect to the $\vec y$ one caused by a larger pressure in the $\vec x$ 
direction due the elliptical shape.
The net effect in terms of the difference of the particle yields between the two directions is larger if the spectra
are decreasing exponentially respect to the case in which they are nearly flat as a function of $p_T$.

\begin{figure}[t!]
\begin{center}
\includegraphics[width=8.3cm]{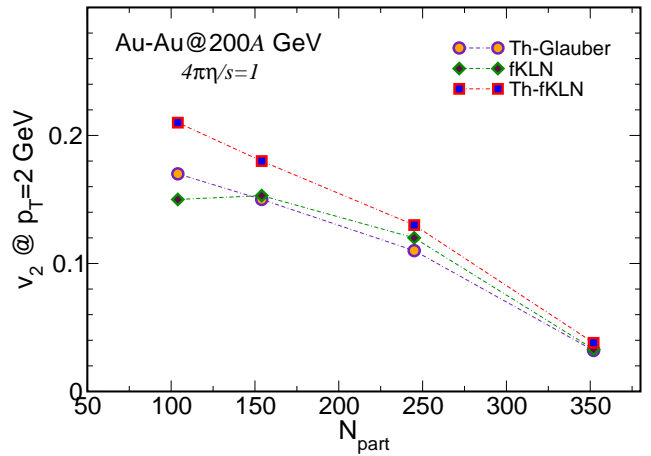}
\caption{Elliptic flow $v_2(p_T)$ at midrapidity $|y|<0.5$ and at $p_T=2$ GeV
for different initial conditions, computed at $4\pi\eta/s=1$.
All the calculations refer to Au-Au collisions
at $\sqrt{s} = 200$ GeV. }
\label{Fig:v2riass}
\end{center}
\end{figure}

In Fig.~\ref{Fig:v2riass} we plot the differential elliptic flow 
at $p_T=2$ GeV for different initializations, as a function of the number
of participants. We have shown results for $b=2.5$ fm corresponding to $N_{part}=352$,
$b=5.3$ fm corresponding to $N_{part}=245$, $b=7.5$ fm corresponding to $N_{part}=154$
and $b=9$ fm corresponding to $N_{part}=104$. This comparison is meaningful since it
permits to visualize and summarize the dependence of $v_2(p_T)$
on the centrality class, comparing the impact of the 
initial distribution on the final $v_2$. In particular the discrepancy between
Th-fKLN and fKLN initializations becomes less relevant for more central collisions,
implying that the effect of the initial momentum distribution is not negligible
if one considers non-central collisions.  
We also note that for central collisions at RHIC Th-Glauber and Th-fKLN for $4\pi\eta/s=1$
predict the same  $v_2$ and the effect is of KLN generating larger $v_2$ disappears.
This is seen also in viscous hydro simulation and it is a further confirmation that our approach
converge to viscous hydro  when the same thermal initial conditions are employed.

\begin{figure}[t!]
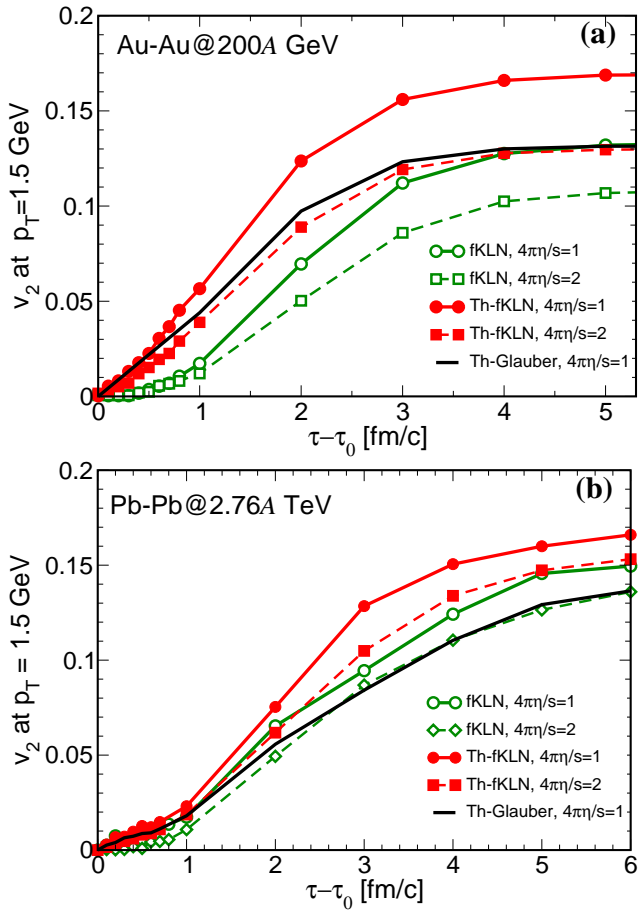

\begin{center}
\includegraphics[width=8.3 cm]{figures/v2_time.eps}\\
\includegraphics[width=8.3 cm]{figures/v2timeLHC.eps}
\caption{(Color online) Evolution of $v_2$ at $p_T=1.5 \rm \, GeV$ as a function of the evolution time 
for all the different initial conditions and $\eta/s$ values.
Upper panel refers to Au-Au collisions
at $\sqrt{s} = 200$ GeV, with an impact parameter $b=7.5$ fm.
Lowere panel corresponds to Pb-Pb collisions at $\sqrt{s} = 2.76$ TeV,
with an impact parameter $b=7.5$ fm.}
\label{Fig:v2_time} 
\end{center}
\end{figure}

In Fig.~\ref{Fig:v2_time}, we plot the elliptic flow as a function of time
for particles at $p_T=1.5$ GeV.
Both the cases of the RHIC and LHC runs are shown in the figure,
for noncentral collisions with $b=7.5$ fm.
In the figure we plot quantities as a function of $\tau - \tau_0$
rather than $\tau$, where $\tau_0$ is the time at which dynamical evolution begins,
because anisotropy is formed thanks to particle interactions which are
effective at $\tau \geq \tau_0$.
In the case of RHIC the behavior clearly splits into two classes: 
one is represented by the thermal distributions;
the other one is given by the case in which the initial spectrum is the fKLN one.
In the case of LHC this distinction is quantitatively less relevant but still present,
probably because as already said the difference between the thermal distribution and the
KLN one is less pronounced.

The main point which we emphasize is that $d v_2/d\tau$ 
for the thermal distributions is large from the very beginning of the dynamical evolution, 
while it is quite reduced for fKLN. This is a further clue that the initial
distribution is very important for the build-up of $v_2$, a statement that we have already
supported discussing the evolution of eccentricities and momentum anisotropy.
We further note that for $\tau-\tau_0 \geq 1\, \rm fm/c$, which is the time range 
roughly corresponding to the thermalization time for all the cases considered here,
$d v_2/d\tau$ of fKLN becomes very similar to Th-fKLN and Th-Glauber. 
This observation further confirms that it is the initial
out-of-equilibrium and nearly saturated distribution that dampens the efficiency in converting the 
space eccentricity into the $v_2$. Therefore, even if thermalization sets in quickly,
as assumed in hydrodynamics $\tau_{isotr} \sim 0.8 \rm \, fm/c$, the actual evolution
from the initial state to the thermalized one cannot be neglected 
in studying the elliptic flow, especially at RHIC energy were most of the anisotropy in 
momentum space develops in the first 3 fm/c at least for semi peripheral collisions.

\begin{figure}[t!]
\begin{center}
\includegraphics[width=8.3cm]{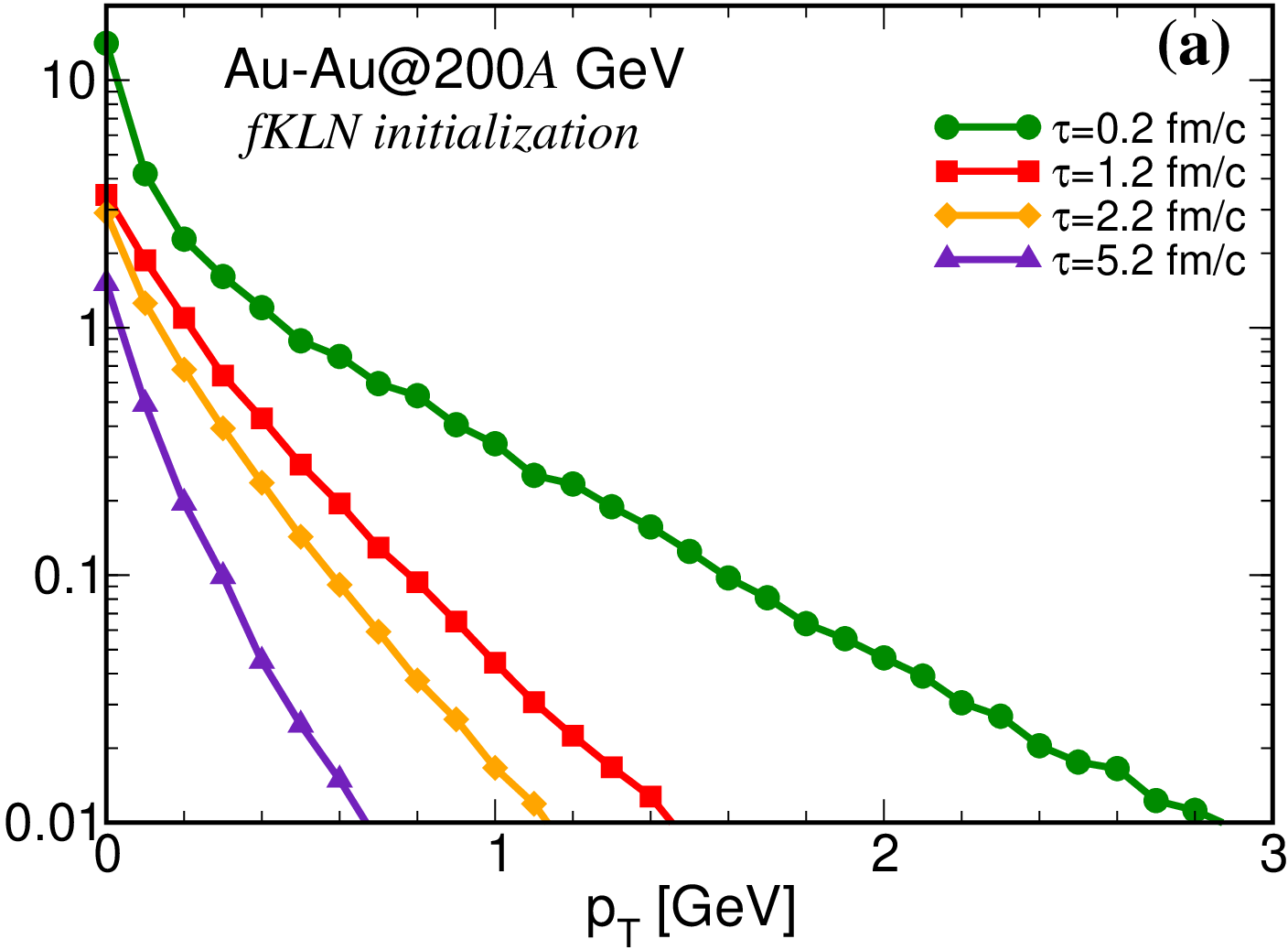}\\
\includegraphics[width=8.3cm]{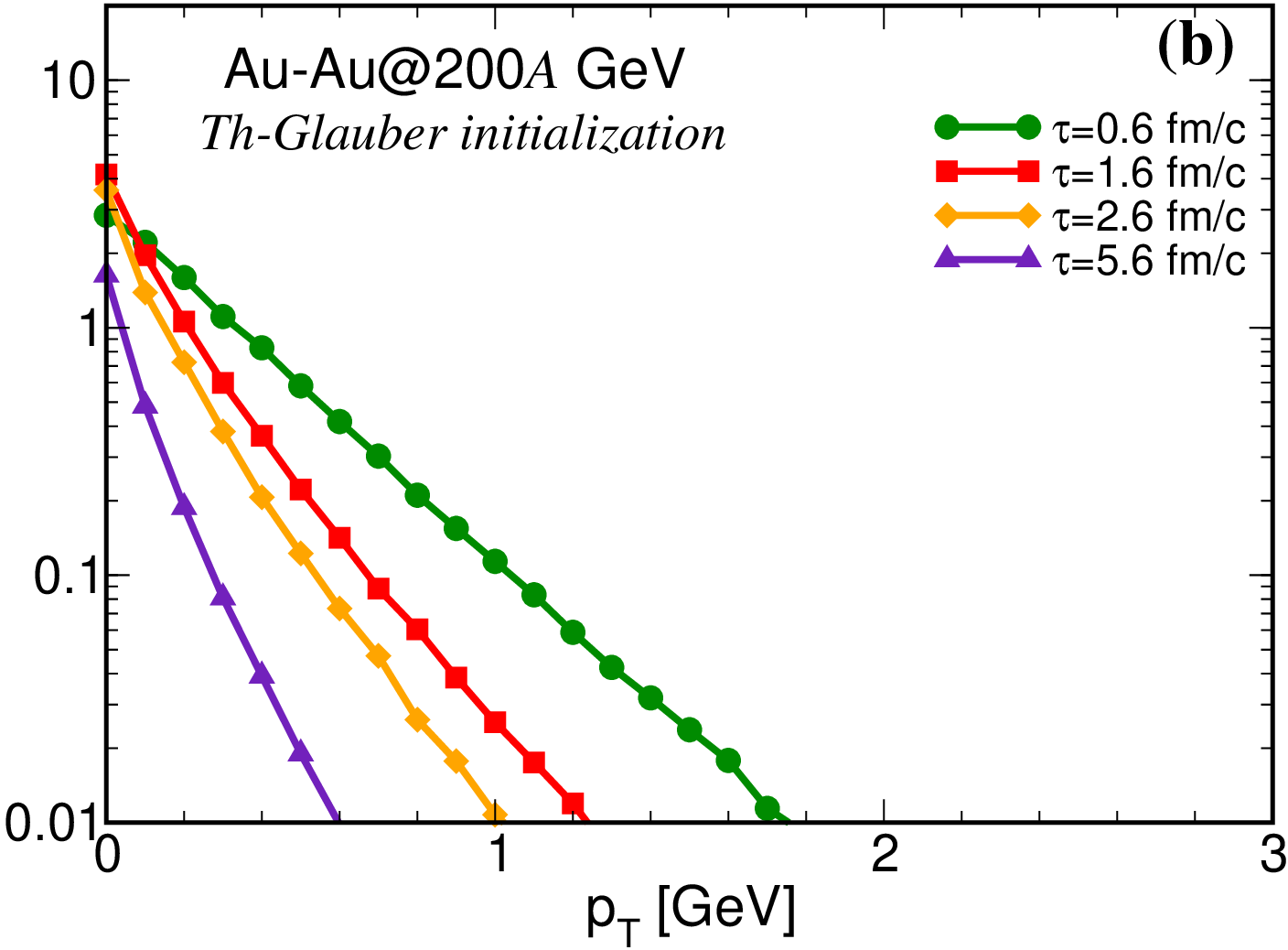}
\caption{(Color online) {\em Upper panel:} Invariant distribution function for several
times, for the case of the fKLN initial condition. {\em Lower panel:}
Same quantities computed for the Glauber initial condition.
All the calculations refer to Au-Au collisions
at $\sqrt{s} = 200 A$ GeV, with an impact parameter $b=7.5$ fm and $4\pi\eta/s=1$.}
\label{Fig:fk}
\end{center}
\end{figure}
\begin{figure}[t!]
\begin{center}
\includegraphics[width=8.3cm]{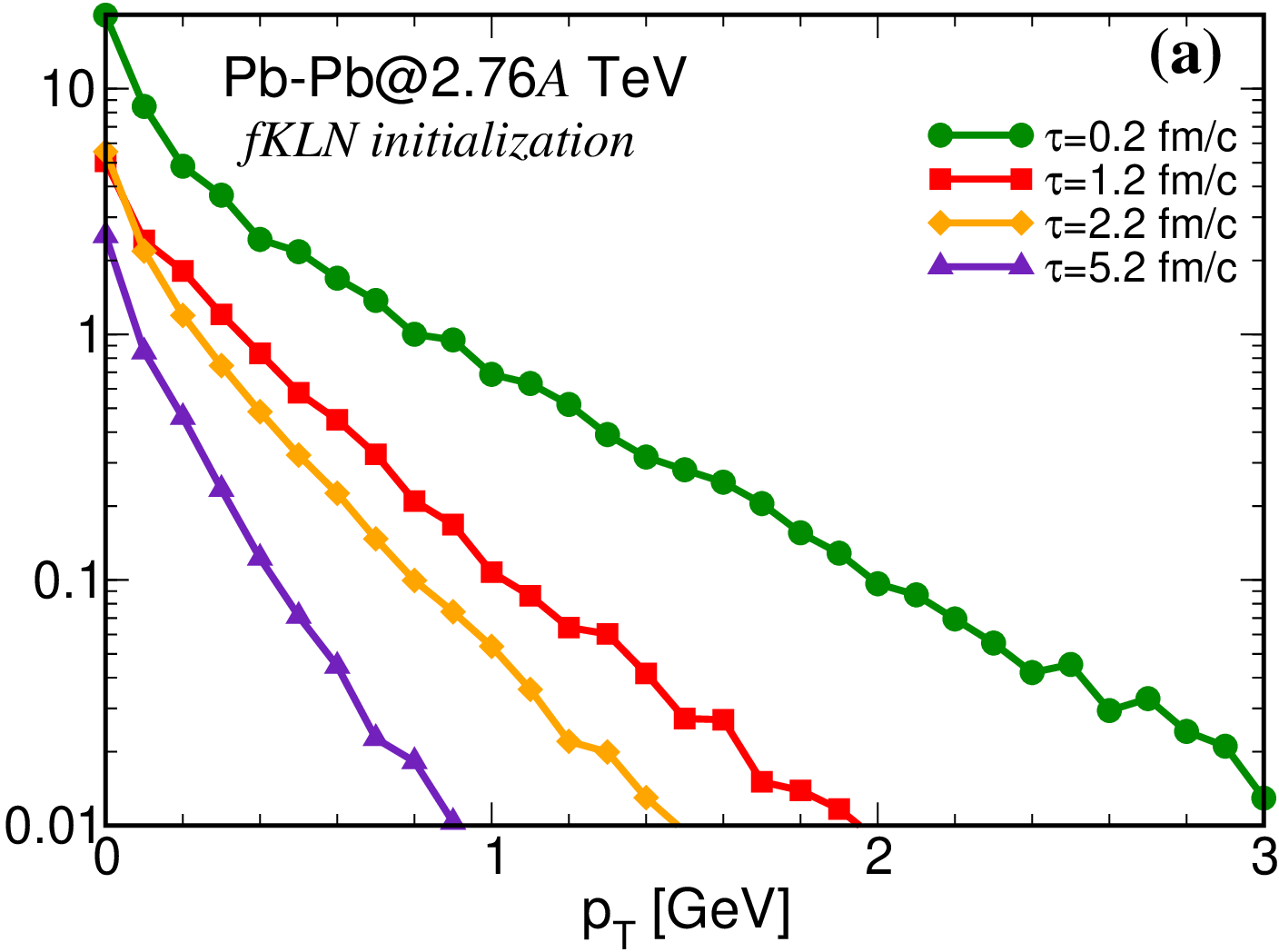}\\
\includegraphics[width=8.3cm]{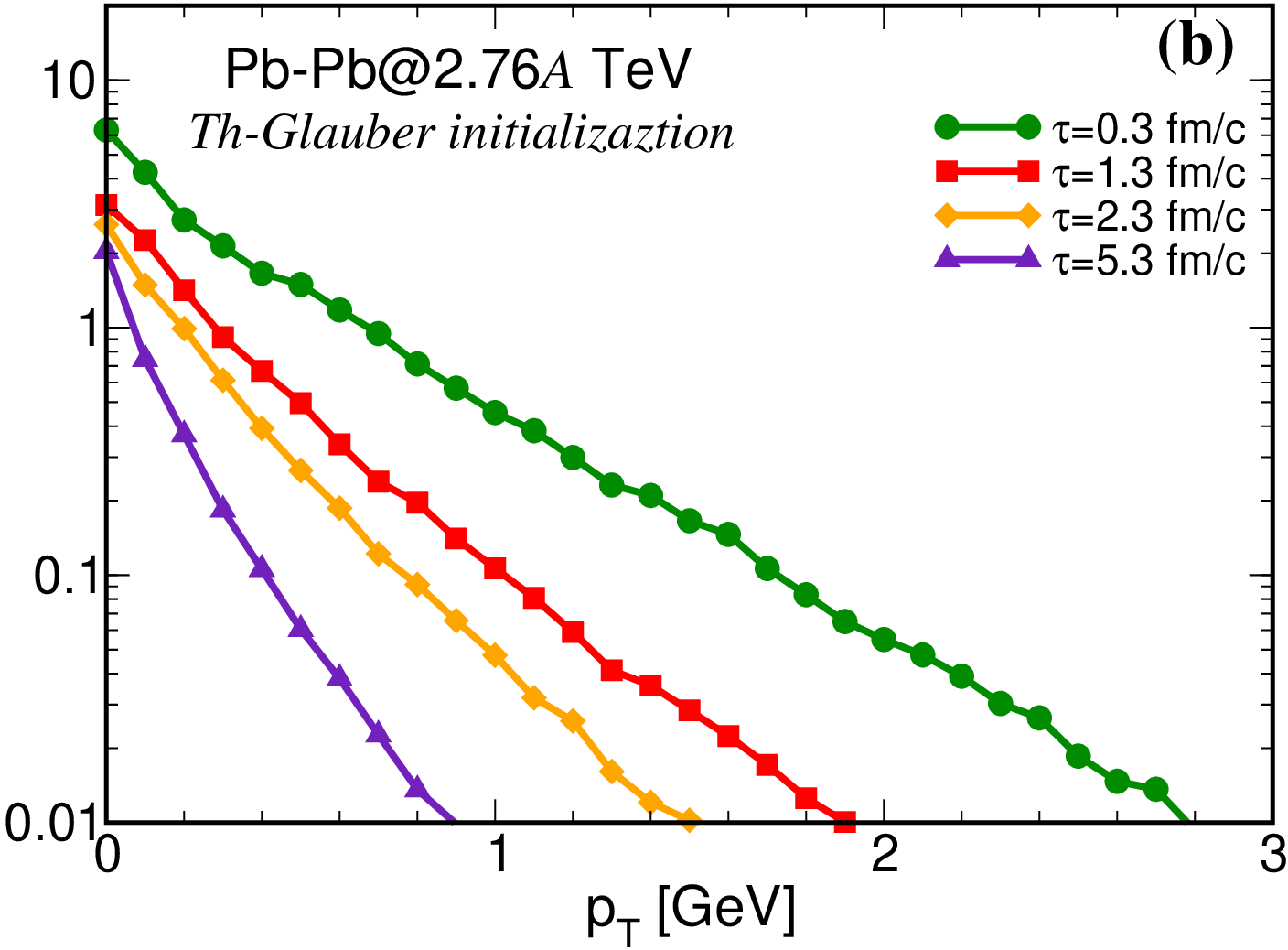}
\caption{(color online) {\em Upper panel:} Invariant distribution function for several
times, for the case of the fKLN initial condition. {\em Lower panel:}
Same quantities computed for the Glauber initial condition.
All the calculations refer to Pb-Pb collisions
at $\sqrt{s} = 2.76 A$ TeV, with an impact parameter $b=7.5$ fm and $4\pi\eta/s=1$.}
\label{Fig:fkL}
\end{center}
\end{figure}

\section{Invariant distribution functions}

Before going to the conclusions we want to discuss the range of values covered by the phase
space distribution functions. This is an issue that recently has attracted a particular
interest because it could be possible that with the high density reached in the
initial stage of the collision the bosonic quantum nature of gluons triggers a
Bose-Einstein condensation~\cite{Blaizot:2013lga}.
The invariant distribution function, $f$, is defined as
\begin{equation}
f = \frac{(2\pi)^3}{g}\frac{\Delta N}{\Delta^2 \bm x_\perp\Delta^2\bm p_T}
\frac{1}{\Delta z \Delta p_z}~,
\label{Eq:fk_def}
\end{equation}
where $N$ counts the number of particles in the phase space volume given by
$\Delta^2 \bm x_\perp\Delta^2\bm p_T \Delta z \Delta p_z$ and $g=8\times 2$ corresponds
to the number of gluonic degrees of freedom. On our grid we have $\Delta^2 \bm x_\perp = \delta_{cell}^2=0.4 \rm\, fm^2$
and $\Delta z = \tau\eta_{cell}$; moreover, we take 
$\Delta^2\bm p_T =p_T\Delta p_T\Delta\phi$ and 
we sample the momentum space
with $\Delta p_T = 0.1~\text{GeV}$. 
We integrate
over momentum rapidity: this is necessary since at initial time $\tau=\tau_0$ 
we assume $y=\eta$, which implies the initial distribution to be singular, namely
\begin{equation}
f = \tilde{f}_0\delta(y-\eta)~,~~~~~\tau=\tau_0;
\label{Eq:f0}
\end{equation}
the subscript $0$ reminds that the above equation is valid only at $\tau=\tau_0$. 
Because of the delta-function the measurable quantity at $\tau=\tau_0$ is $\tilde{f}_0$,
which can be obtained by integrating $f$ over momentum rapidity; 
when we compute the distribution function at later times, for consistency we also
integrate over momentum rapidity. For later convenience
we also integrate over the azimuthal angle $\phi$. We then define
\begin{equation}
\langle f\rangle=\frac{1}{2\pi}\int d\phi\int f  dy~,
\label{Eq:def_ave}
\end{equation}
which on each cell of the configuration space reads
\begin{equation}
\langle f\rangle=\frac{1}{2\pi\tau}\frac{1}{\eta_{cell}\delta_{cell}^2}
\sum_{cell}
\frac{1}{E p_T\Delta p_T}~,
\label{Eq:def_ave_grid}
\end{equation}
with $E=\sqrt{p_z^2 + \bm p_T^2} = p_T\sqrt{1+\sinh^2y}$ 
and the sum is understood over all the particles that at time $\tau$
are in the specified cell.

Here we present the evolution of $f(p_T)$ in the central region
of a QGP fireball to have a first estimate of the value explored with time
and transverse momentum $p_T$. In Figs.~\ref{Fig:fk} and~\ref{Fig:fkL}
we plot the invariant distribution functions for two different initial conditions.
The upper panels of the figures correspond to the fKLN, while the lower panels
correspond to the Glauber initial condition; Fig.~\ref{Fig:fk} is the result
for the Au-Au collisions at the RHIC energy, while Fig.~\ref{Fig:fkL} corresponds
to the case of Pb-Pb collisions at the LHC energy. 
In order to obtain the results shown in the figure, we have integrated Eq.~\eqref{Eq:def_ave_grid}
over the region $|x_T|<2.45$ fm, $|y_T|<4.55$ fm and $|\eta|<0.1$,
which embed the hottest and densest parts of the initial fireball. 
However, the results are qualitatively
similar if we chose to compute $\langle f\rangle$ by means of Eq.~\eqref{Eq:def_ave_grid} only in
the central cell.
The natural consequence of the longitudinal (as well as the transverse) expansion is that the distribution function
becomes smaller and smaller as time increases. In fact, from Eq.~\eqref{Eq:def_ave_grid} we
expect $f\sim\tau^{-1}$ due to the longitudinal expansion close to 
luminal velocity. 

In Fig.\ref{Fig:fkL} (upper panel) we can observe that at $p_T \sim \rm 1 \, GeV$ the initial $\langle f(p_T) \rangle \sim \, 0.5$ 
but the rapid longitudinal expansion along with the rapid cooling down is such the $f$ decreases very 
quickly with and at $\tau \sim 1 \rm\, fm/c$ it is already smaller than $5 \cdot 10^{-2}$. On the hand at
$p_T<0.5\rm \, GeV$ not only at $\tau_0$ the $\langle f(p_T)\rangle >1$ but due to the high scattering 
rates that drive a rapid thermalization the decrease of $\langle f(p_T)\rangle $ with $\tau^{-1}$ is fully damped
at even at $\tau \sim \, 1-2 \rm\, fm/c$ we have $\langle f \rangle \sim 1$.
A similar trend with even larger value of $\langle f \rangle$ is observed at LHC and not only for the fKLN
but also for the Th-Glauber initial conditions.

We remind that in our formulation we have not introduced the bosonic enhancement factors in the collision 
integral, which however are expected to give a relevant effect if $f\gtrsim1$.
Still our result shows that the longitudinal expansion coupled to the thermalization process keeps
the occupation number quite large for a relatively long time interval and therefore it is reasonable to expect that
the formation of a transient Bose-Einstein condensate also in the expand glasma in HIC and not only
for the static case of a system in a box \cite{Blaizot:2013lga}.
The results of~\cite{Blaizot:2013lga}
are of course quite interesting and we plan to improve our calculation in the near future 
to take into account the Bose enhancement factors in the collision integral,
as well as the inelastic processes which convert gluons to quarks. 

As for the effect discussed on the build-up of $v_2$ in Section IV, we do not expect significant impact
of the Bose-Einstein enhancement factors $1+f$ in the collisions integral because, as discussed,
only for small $p_T$, $f$ is relatively large and furthermore only in the very central region of the fireball. 
while our effect take place for $p_T> 1 \rm\, GeV$. 
Moreover, the effect we find is mainly due to the initial non-equilibrium dynamics and
not to the fixed point of the kinetic equation which instead determines the
final state. Further investigations are in progress.

\section{Conclusions}
In this article we have presented our results on thermalization,
isotropization and building-up of the elliptic flow 
for fireballs produced in relativistic heavy ion collisions both at 
RHIC and LHC energies. We have put emphasis on the role of a nonequilibrium initial
condition on the generation of the collective flow, using as a model the fKLN-glasma
initial condition which differs from the thermal distribution for the presence of a saturation scale
in the low $p_T$ spectrum. Our study is based on kinetic theory at fixed $\eta/s$.

In our study we have neglected the initial time evolution of the
chromo-electric and chromo-magnetic fields (the glasma) produced immediately after the
collision, whose intensity can be directly related to the saturation scale $Q_s$
in the CGC before the collision. 
Therefore our approach could be justified as soon as the initial strong glasma longitudinal fields decay 
into particle quanta and its longitudinal pressure becomes positive.
The characteristic time for this decay 
is of the order of $\tau_0\approx 1/Q_s$~\cite{Ryblewski:2013eja,Gelis:2013rba,Fukushima:2013dma}. 
The advantage
to use kinetic theory from $\tau_0$ rather than hydro is that the nonequilibrium
initial distribution is not problematic as kinetic theory is built 
to study the evolution of a generic $f(x,p)$ distribution function. 

As a non-equilibrium initial distribution in our simulations we have used the 
spectrum obtained within the fKLN model,
which embeds a saturation scale in momentum space. In both RHIC and LHC runs we have found that
thermalization times are $\Delta\tau_{therm} \sim$ 0.8-1 fm/c. 
Moreover we have studied the time scales for the isotropization of the expanding fireball,
defining transverse and longitudinal pressures, $P_T$ and $P_L$ respectively,
and computing the time evolution of $P_L/P_T$. We have found that isotropization
is not reached if  $\eta/s \geq 0.3$, in agreement 
with~\cite{Ryblewski:2013eja,Gelis:2013rba}; this result is easily
understood, since higher viscosity implies lower cross sections
among partons and dynamics in the longitudinal direction
and transverse plane are decoupled. On the other hand it is significant that
for a fluid with $\eta/s\sim 0.1$ also isotropization ($P_L/P_T \simeq 0.7$)
occurs very quickly $\tau \sim 0.5\rm\, fm/c$.

We have then focused our attention on the elliptic flow production
when the initial distribution has a saturation scale built in it.
We have found that the amount of elliptic flow produced in heavy ion collisions
depends not only on the pressure gradients and the $\eta/s$ of the system, 
but also on the initial distribution in momentum space. 
In particular, an initial condition characterized by a momentum distribution 
with a saturation scale generates smaller $v_2$ respect to the thermal one.
This result is quite general, and we expect it should be valid, besides QGP in uRHICs, 
for systems like cold atoms in a magnetic trap which are characterized by a value of $\eta/s$ 
close to the QGP one~\cite{O'Hara:2002zz}.
Assuming the fKLN distribution as the one arising from the melting of the glasma,
the effect of the initial nonequilibrium distribution affects the estimate of $\eta/s$ 
of about a factor of two. 
However we have seen also that this effect is maximal for semi-peripheral collisions,
becoming quite small for very central collisions.

The relevance of our results is further enhanced by the fact that Th-fKLN with 
$4\pi\eta/s \sim 2$ would generate a low $v_3$ respect to the available data.
This is the main conclusion of~\cite{Adare:2011tg}, namely that Th-fKLN is not able to describe
for the experimental observations, might be revised.

In order to make more precise comparison with 
experimental data we are currently implementing also fluctuating initial conditions 
which will allow to extend the present study to all the $v_n$ harmonics relevant in HIC's.
This will allow to see if fKLN can account for the measured $v_3$ or the non-equilibrium 
damps such harmonics even more than what seen in hydrodynamics. 
In~\cite{Gale:2012rq} it is found that harmonics up to the fifth order
can be reproduced by combining CYM early-time with hydro late-evolution
evolutions; however in the calculations of~\cite{Gale:2012rq} 
a small deviation from equilibrium has to be assumed in order to 
use viscous hydrodynamical equations. Therefore it will be interesting to compute
the higher order harmonics combining the CYM initial spectrum 
with the dynamics embedded in the kinetic equations where the assumption
of sudden thermalization can be relaxed.

Finally, we have also computed the invariant gluon distribution functions, $f_g$, for
the fKLN glasma and Glauber initial conditions. We have found that the 
initial longitudinal expansion strongly affects the evolution of these
distributions at early times, causing their lowering of one order of magnitude
within 1 fm/c for $p_T\gtrsim 1$ GeV. 
On then other hand at $p_T\lesssim 0.5 \rm\, GeV$ the longitudinal expansion is compensated by
the fast thermalization and even without including $(1+f)$ enhancement factors $\langle f \rangle $
stays larger than unity for a time interval $\Delta \tau \sim 2 \rm\, fm/c$. this suggested that even
in the fast expanding QGP in HIC a transient Bose-Einstein condensation can take place
as suggested in  ~\cite{Blaizot:2013lga,Blaizot:2012qd,Huang:2014iwa}. 
However the effect of Bose statistics should be relevant at $p_T\sim T$,
while our effect is related to the initial non-equilibrium and not
to the final exact distribution.
In order to verify quantitatively
how this attractor works in presence of the longitudinal expansion we have to
perform simulations with Bose enhancement factors in the collision integral kernel,
which is beyond the scope of the present study but that we will perform 
in the near future. 

{\em Acknowledgements.}
The authors acknowledge enlightening discussions with M. Chernodub, J. Y. Ollitrault
and N. Su. M.~R. acknowledges Mei Huang for her kind hospitality at Institute
of High Energy Physics in Beijing, where part of this work was completed.
V.~G. and F. S. acknowledge the ERC-StG funding under the QGPDyn grant.

\end{document}